\documentclass[12pt]{article} 
\input epsf.tex
\usepackage[dvips]{graphicx}
\usepackage{epsfig}

\def\m2 {m$^{2}$ }
\def\degree {$^{\circ}$ }
\def\deg1 {$^{\circ}$}

\def\t10 {$t_{10}$ }
\def\t50 {$t_{50}$ }
\def\t90 {$t_{90}$ }

\begin{document}
\begin{titlepage}
\begin{center}

{\bf \Huge TANGO ARRAY}
\vskip 0.5 cm
{\bf \Large An Air Shower Experiment in Buenos Aires$^{\dagger}$}
\vskip 0.8 cm
{\large P.~Bauleo, C.~Bonifazi, A.~Filevich$^{1}$ and A.~Reguera$^{2}$}

{\em \small Departamento de F\'{\i}sica, Comisi\'on Nacional de Energ\'{\i}a 
At\'omica,\\ Avenida~del~Libertador~8250, (1429) Buenos Aires, Argentina}

\vskip 1.5cm
{\bf Abstract}
\vskip 1.5cm
\end{center}

{A new Air Shower Observatory has been constructed in Buenos Aires during 1999, 
and commissioned and set in operation in 2000. The observatory consists of 
an array of four water \v{C}erenkov detectors, enclosing a geometrical area of $\sim$~30.000~m$^{2}$,
 and is optimized for the observation of cosmic rays in the ``knee'' energy 
region. The array detects ~$\sim$~250 to~$\sim$~1500 showers/day, depending 
on the selected triggering condition. In this paper, the 
design and construction of the array, and the automatic system for data 
adquisition, daily calibration, and monitoring, are described.
Also, the Monte Carlo simulations performed to develop a 
shower database, as well as the studies performed using the database 
to estimate the response and the angular and energy resolutions of 
the array, are presented in detail.}

\vfill  

\footnotesize{$^\dagger$ Further information available in {\bf http://www.tango-array.org}

	      $^1$  Fellow of the CONICET, Argentina.

	      $^2$  Now at Telesoft S.p.A., Buenos Aires.}

\end{titlepage}

\section{Introduction}
The Earth's atmosphere is being bombarded continuously by a flux of particles 
(cosmic rays), coming from all directions. Their energies range from a few 
MeV to more than 10$^{20}$~eV. Their spectrum follows a power law with a 
negative exponent which is almost constant over thirteen orders of magnitude in 
energy. The origin of the cosmic rays is still an open question. Those rays with 
energy below $\sim$ 1 GeV are likely to have a solar origin, but for higher 
energies their acceleration mechanism remains in mystery. It is believed that up 
to $\sim$~4.10$^{15}$~eV they can be accelerated by diffuse shock processes 
produced in supernova explosions. In this energy region, (usually called ``the 
knee''), the exponent of the power law describing the cosmic ray flux per units 
of area, time, solid angle, and energy, suddenly steepens from $\sim$~-2.7 to 
$\sim$~-3.2, and this change is believed to be related with the maximum 
energy that can be transferred by a supernova shock to a single particle. If the 
kinetic energy of the cosmic ray is high enough, then secondary particles are 
produced as a consequence of hadronic or electromagnetic interactions with 
the upper atmosphere atomic nuclei. Those secondary particles will, in turn, 
produce more particles, yielding a cascade which is known as an Extensive Air 
Shower (EAS). Depending on the primary energy and zenithal angle, this cascade 
can be stopped in the atmosphere, or even reach the ground level. 

A method which has been used for the observation of EAS is the detection of the 
light emitted by the \v{C}erenkov effect in air, while fast charged particles, 
(mainly electrons), are crossing the atmosphere (WHIPPLE, CANGAROO). 
Alternatively, it is possible to observe the UV light emitted by decay processes 
occurring in the atmospheric molecules after excitation by the EAS's secondary 
particles (Fly's Eye, HiRes, Pierre Auger Project, Telescope Array, OWL Project).
The amount of UV and \v{C}erenkov light emitted by an EAS is extremely 
faint, and because of this it is possible to observe these processes only 
during moonless dark nights, and by using relatively large telescope mirrors as 
light concentrators and sensitive photomultiplier tubes. 

Another (and perhaps more common) approach (Haverah Park, AGASA, Volcano Ranch,
SUGAR) is the direct detection of the shower secondary 
particles reaching the ground level. The size of the footprint at ground 
level is several thousand square meters for showers produced by primary cosmic 
rays of energies near the ``knee'' or higher. Because of this these experiments 
are designed so as to observe only samples of the particle showers using an array of ground-based 
detector stations, where gas-filled chambers, plastic scintillators or 
\v{C}erenkov-effect detectors are typical components.

The detector stations of these ground arrays are usually capable to measure 
particle densities. In the case of the array described in the present work, 
where water \v{C}erenkov detectors (WCD) are used, this measurement is performed 
through a sample of the amount of light emitted when the shower particles 
traverse through the water radiator. Also, the precise relative times of the 
signals produced by each station are recorded, together with the \v{C}erenkov 
light intensity 
information. By using the relative hitting times at each station and the known 
geometry of the array it is possible to determine the direction of arrival of 
the primary cosmic ray, assuming that the general development of the shower follows a 
rather flat front profile.  Rigorously the shower front is a curved surface 
whose radius of curvature could in principle be determined if the number 
(and quality) of the sampling detectors is high enough.

The determination of the primary energy from EAS measurements using ground-based 
detectors is closely tied to shower reconstructions, based on Monte Carlo 
simulations. These simulations correlate the primary energy to the particle 
densities at a fixed distance from the shower ``core'' position, that is, the 
center of gravity of the air shower at ground. In a simplified model, the 
primary energy is simply estimated as a magnitude proportional to the total 
number of particles in a shower. Hence, the particle density measurements 
performed by each station is used to estimate the total number of particles of 
the shower.

In the following sections the design and construction of this new air shower 
experiment, which has been optimized for the ``knee'' region of the energy 
spectrum, are described. The necessary simulations, which were required to set 
the numerous design parameters of the array, are presented in detail. 

\section {The array}

\begin{figure}[h]

\centerline{\epsfig{file=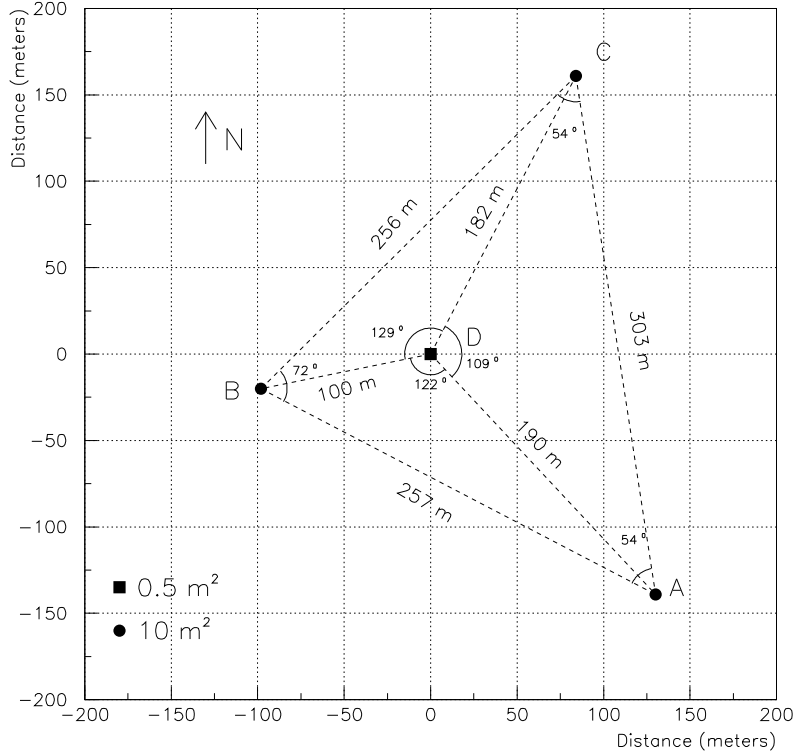,width=11.5cm,height=11.5cm }}

\caption{TANGO Array Layout. Circles indicate the positions of the three 
10~m$^{2}$ stations, whereas the square shows the position of  the central, 
0.5~m$^{2}$ detector. The distances shown in the figure have been measured using 
a GPS.}
\label{fig:array}
\end{figure}

The TANGO ({\bf TAN}dar {\bf G}round {\bf O}bservatory) Array has been 
constructed in Buenos Aires, Argentina, at ($\sim$ 15 m a.s.l), 35\degree 34' 
21''~S and 58\degree 30' 50''~W, in the Campus of the Constituyentes Atomic 
Center, belonging to the Argentinean Atomic Energy Commission (CNEA). The data 
acquisition (DAQ) room was set inside the TANDAR Accelerator Building. Three 
detectors are placed on the vertices of an almost isosceles triangle, and a 
fourth detector was installed on top of the building, in a convenient position 
close to the center of the triangle, as shown in Figure \ref{fig:array}. The final 
positions 
of the detector stations were conditioned by the free space available between 
the existing buildings, and an effort was made to come up to an overall shape as 
close as possible to an equilateral triangle, which maximizes the effective 
collection area. The distances between surface stations were measured using a
GPS and their error has been estimated in $\pm$~1~m (the measurement was performed 
after release of the high precision GPS service). The final configuration 
encloses a geometrical area of 31286~m$^{2}$. The array has a yearly average 
overburden of $\sim$~1000~gr/cm$^{2}$.

During an EAS event the DAQ system measures both, the intensities of the 
\v{C}erenkov photons 
emitted by the water when crossed by the secondary particles of a high energy 
cosmic ray's EAS, and also, the arrival time of the these particles to each 
station. The threshold energy of the array, resulting 
from the geometry and from the particular detector conditions, (present noise, 
trigger levels, etc) 
is close to 10$^{14}$ eV for vertical showers. The detector stations are connected by low 
attenuation (RG-213) coaxial cables to the DAQ room, where the signals are 
recorded using a 4-channel digital oscilloscope connected to a computer. 
Depending on the selected trigger conditions, which are generated by standard 
NIM electronic modules, the number of accepted events ranges from $\sim$~250 
to $\sim$~1500 per day.

\subsection{The detector stations}

\begin{figure}[h]
\centerline{\epsfig{figure=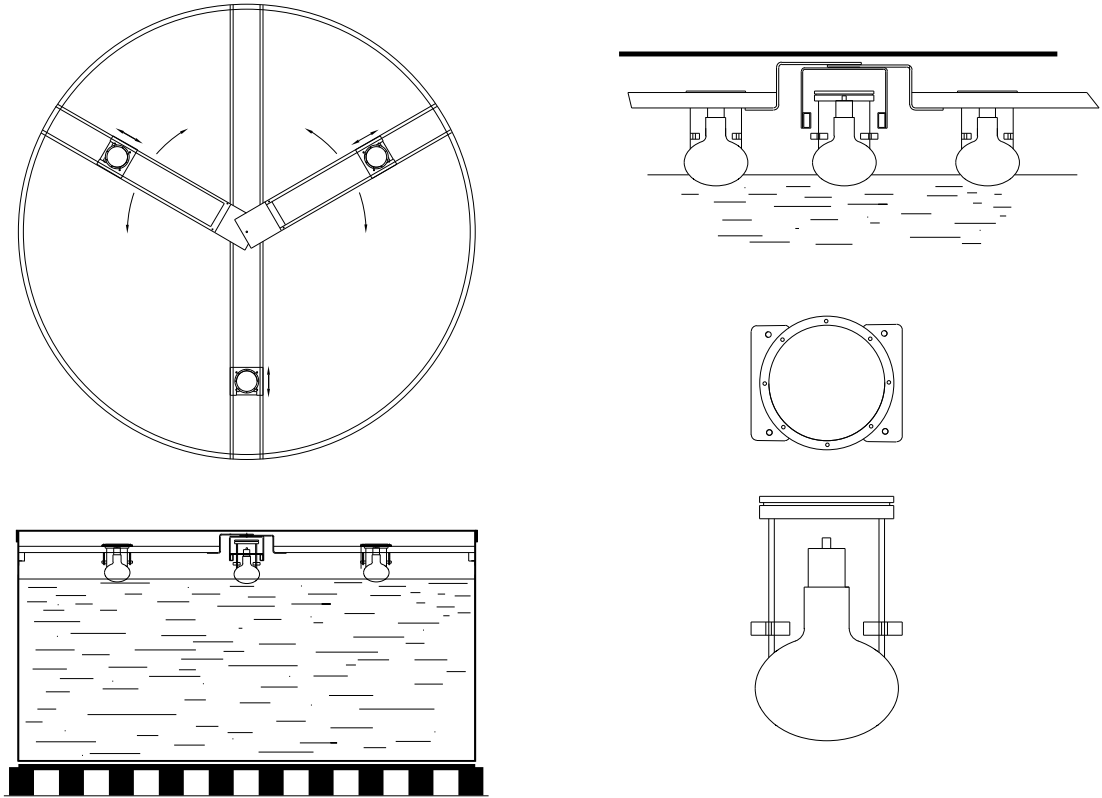,width=16.5 cm,height=11.0 cm}}
\caption{Top and side view of the first 10~m$^{2}$ detector used in the experiment.
In the right is shown a detail of the PMT's enclosure.}

\label{fig:detector}
\end{figure}

This array is a project which grew up from the first 1:1 scale prototype of a 
WCD \cite{nimtank}(See Figure \ref{fig:detector}) built in 1995 by members of 
the local Pierre Auger Project 
Collaboration \cite{PAP}. This first detector (labelled A in Figure \ref{fig:array}) was
construced in a tank, cylindrical in shape, made of 0.68~mm stainless steel plate, with
a footprint area of 10~m$^{2}$. The effective water depth is 120~cm. Three, 8-inch 
photomultipliers (Hamamatsu R1408), symmetrically placed at 120~cm from the 
tank axis were installed looking down on the top of the detector, having only 
the photocathodes immersed in the water working as \v{C}erenkov radiator. Thus, a 
sample of the \v{C}erenkov photons emitted when a charged particle crosses the 
tank are collected by the three PMTs. This detector, being a prototype, was 
designed as a flexible system, allowing the introduction of modifications in the 
photomultiplier positions, the effective water height, or the inner lining 
material. During the measurements as a component of the TANGO array the 
configuration of this detector was that of the Pierre Auger Project baseline 
design\cite{PAP}, with the dimensions mentioned above. In order to improve the optical 
properties of the inner surfaces all detectors were fully lined with Tyvek which 
is a highly UV-diffusive and reflective material\cite{tyvek}.

\begin{figure}[h]

\centerline{\epsfig{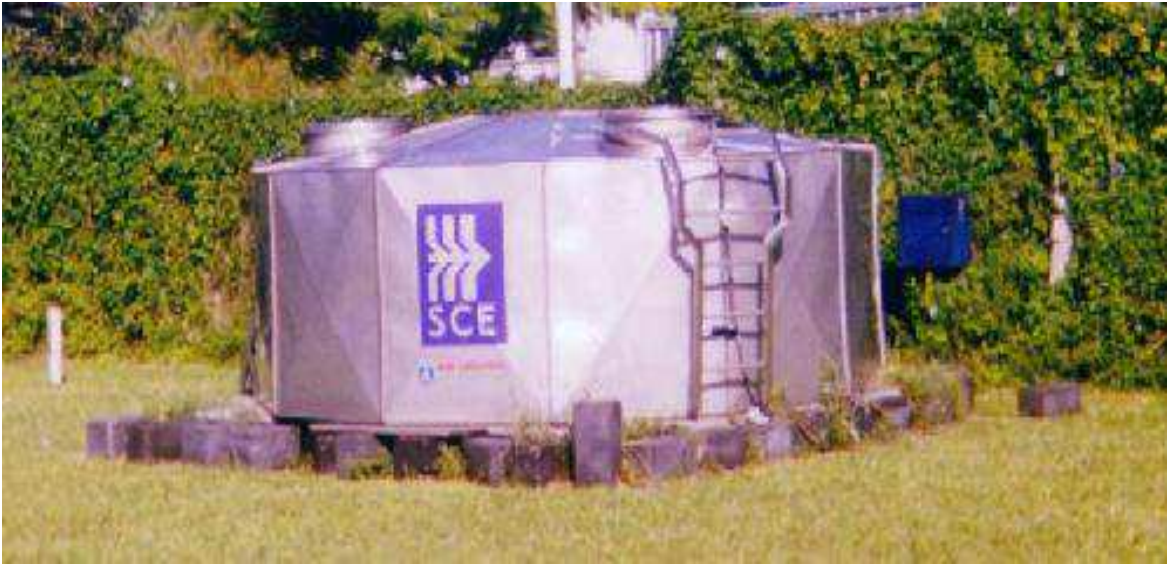}}
\caption{Detector B. The cabinet installed on the wall at the right side contains 
the local electronics and the high voltage power supply.}

\label{fig:station_1}
\end{figure}

The two other detectors sitting in the vertices of the triangle (B and C in 
Figure \ref{fig:array}) have the same general dimensions quoted previously. 
They are made of 1~mm thick stainless steel, and the external walls are shaped as 
a dodecagon (see Figure \ref{fig:station_1}). In these detectors we used Hamamatsu R5912,
 8-inch diameter 
PMTs, arranged with the same geometry used in the first detector tank. The 
fourth detector (D) is smaller, it was made using a fiberglass-reinforced 
polyestyrene tank, with a footprint of 0.5 m$^{2}$ and an effective water depth of 
80 cm. The tank was also internally lined with Tyvek and only one, 3-inch PMT, 
was installed centered on the top of the tank.

The larger outer detectors are more adequate to measure lower particle densities 
generated by showers falling relatively far away from them, either close to the center 
of the array or very away the whole array. The relatively large ratio between the 
volume of the external detectors to that of the smaller central detector helps to improve 
the accuracy in the determination of the particle densities in those cases where 
the shower core falls close to the center of the array. This is so because of the 
larger dynamic range of the central detector, which admits higher particle densities without 
going into saturation (See Section \ref{lab:elect}, and the higher sensitivity of the larger
detectors placed on the vertices of the triangle.

All PMTs used in the WCDs were mounted in water-tight enclosures that protect 
from moisture their voltage dividers and only the photocathode areas of the glass
bulbs are immersed in the water radiator (See Figure \ref{fig:detector}. The glass bulbs 
of the PMTs were glued to the PVC housings using an elastic silicone compound to reduce 
mechanical stresses that could break the glass, as happened in the Milagrito 
experiment\cite{Milagrito}. Local high voltage power suplies, fed from the AC 
mains, were installed near each station. The bias configuration of all PMTs was adopted 
as grounded cathode, to prevent eventual noise produced by electrical leaks or 
discharges through the glass. 

The water used to fill the tanks was treated in a reverse-osmosis plant, 
producing an average final water resistivity of about 1~M$\Omega$-cm. Before 
filling the tanks they were carefully degreased, brushed with water and mild 
detergent and rinsed abundantly with the same water used as the detector 
material. These precautions, together with the darkness and the fact that the water used as 
detector material has a very low level of bacteria nutrients, virtually blocked 
any extensive biological activity \cite{gap_96_036}. After more than 
1 year since the filling of the detectors, no significative decrease in the signal 
strength has been observed.

\begin{figure} 
\epsfig{figure=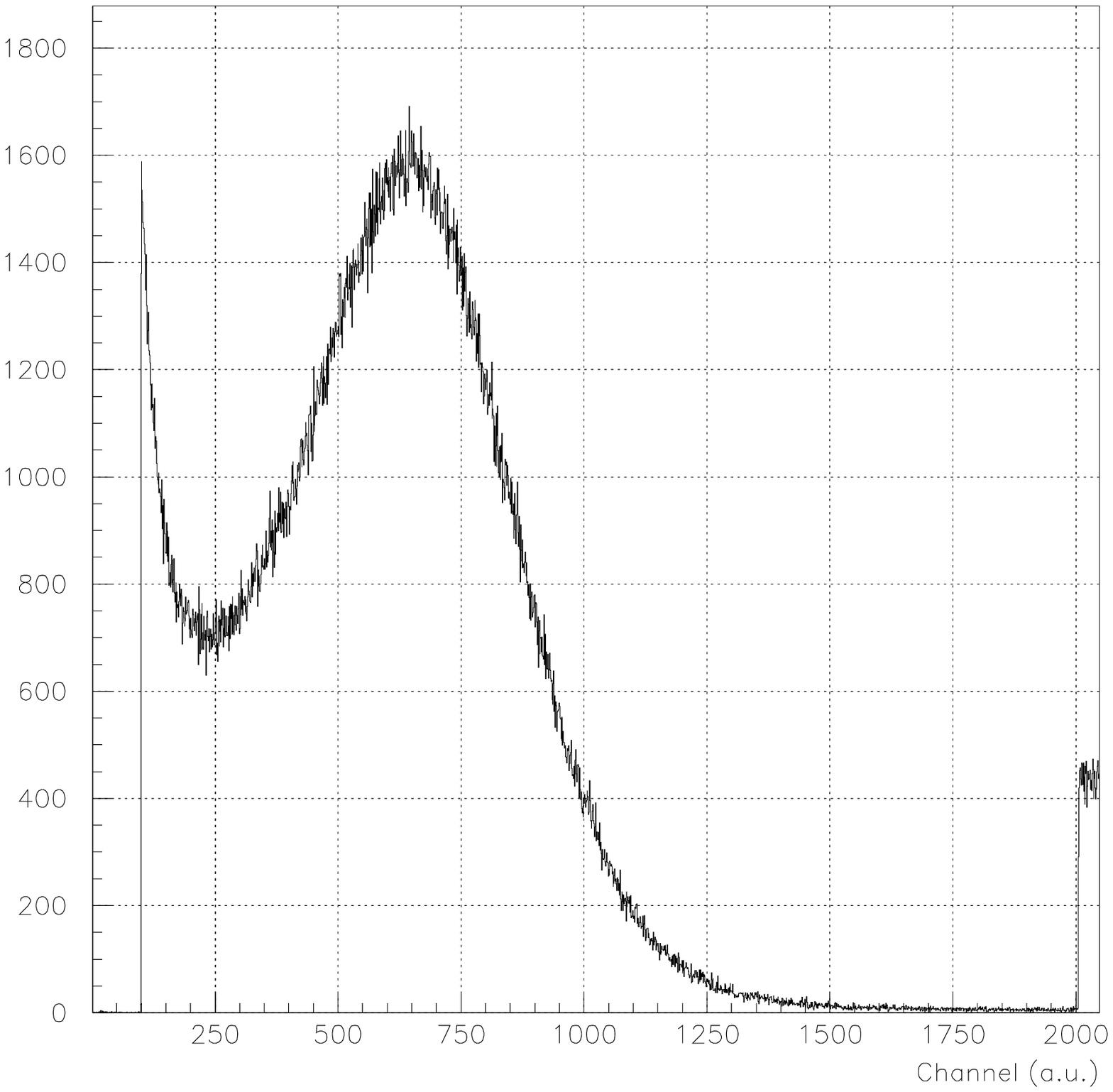,width=7.5 cm,height=7.5 cm}  
\epsfig{figure=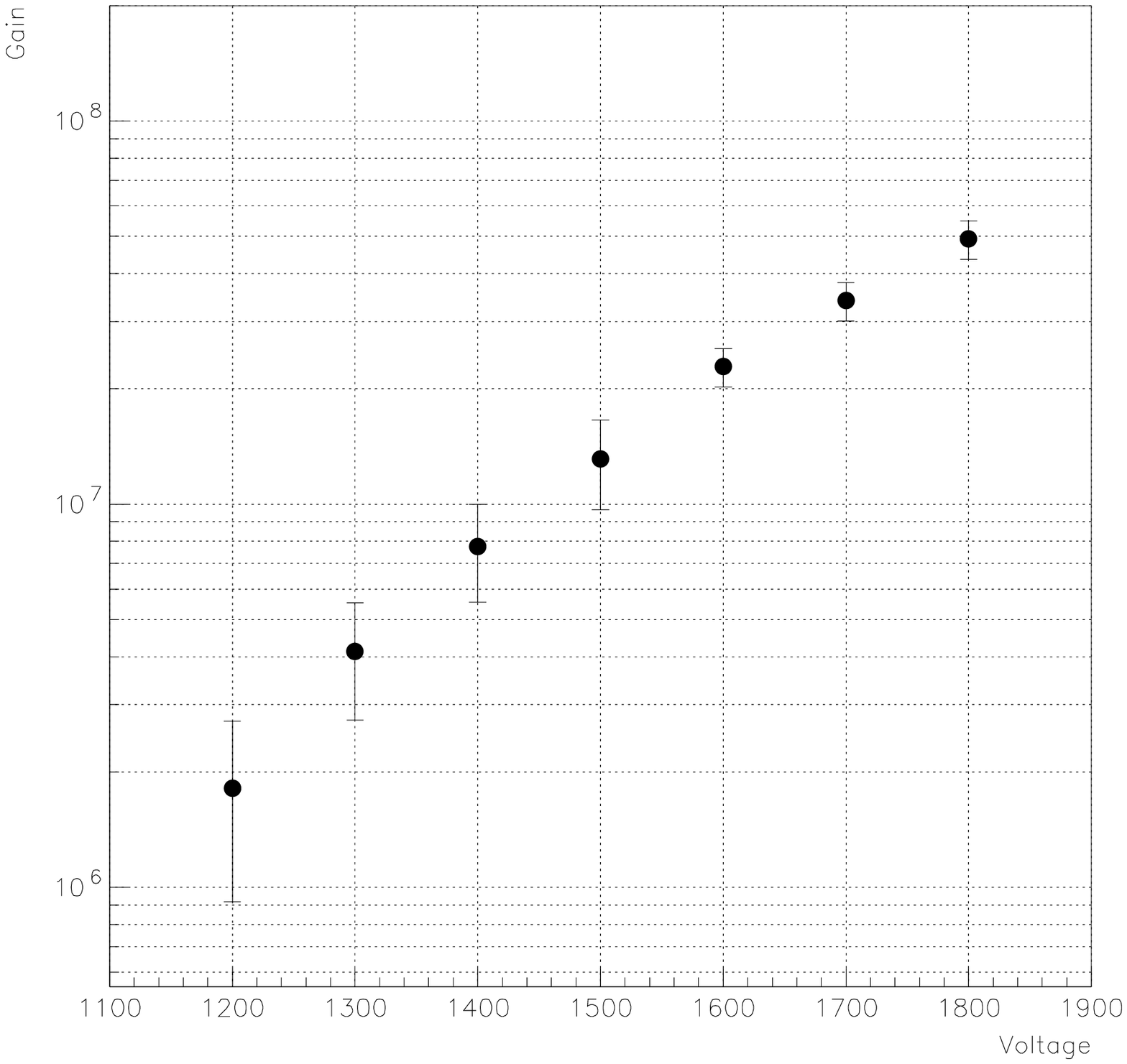,width=7.5 cm,height=7.5cm}
\caption{Typical single electron spectrum measured at 1500 Volts (left) and gain (right)
 for a Hamamatsu R5912 tube measured using the dark box.}
\label{fig:se}
\end{figure}

\subsection{Characterization of the photomultiplier tubes}

The gains of the three (R1408) PMTs used in the first prototype were measured 
previously\cite{gap_ganancia}, and we have built a dark box, adequate for measuring 
the gain and dark current of the new Hamamatsu R5912 tubes, purchased for the new 
detector stations. We also used this darkbox to characterize the photocatode sensitivity 
profiles and the influence of the Earth magnetic field direction of the PMTs. The 
dimensions of the box are 50 x 50 x 100~cm and accepts one PMT, which is mounted 
in an axially rotatable holder. 

Using electrons thermally emitted from the photocathodes at room temperature, we 
measured the gains of the PMTs by means of the single electron technique. In 
these measurements we tested different voltage divider configurations in a range 
of high voltage from 1100 to 1800~V. In all cases the tubes were kept in total 
darkness for at least 2 hours before collecting the single electron spectra, to 
reduce the rate of multi-electron emissions due to fast decaying fluorescence in 
the photocatodes. The dark current pulse rate (threshold = 1/3 p.e.) after one hour of 
storage in total darkness was about 900 Hz at 1500 Volts. A typical spectrum and 
a plot of the gain values are shown in Figure \ref{fig:se}.

A study of the influence on the PMT gain of the Earth magnetic field direction 
relative to the dynode geometry, was also performed. The gains were carefully 
measured at a fixed HV setting (1.5 kV) for 8 different azimuthal orientations 
of the PMTs, covering 360\degree. The tubes were measured keeping always their 
axis in vertical position. No significative shifts (less than $\pm$ 1\%) were observed in 
the peak positions of the single electron spectra obtained in this way. From 
these results we concluded that the local influence of the Earth magnetic field 
direction on the gain can be neglected, and thus no special care was taken to 
install the tubes in the detector stations at any special dynode orientation.

\subsection{Gain settings and calibration of the detector electronics}
\label{lab:elect}

After testing several possible configurations we adopted a pasive, grounded 
cathode design for the voltage dividers, as simple and reliable as possible, 
assuring both wide dynamic range, and linearity. The final configuration chosen 
is similar to that recommended by Hamamatsu for the R5912 tubes. Metal film 
resistors were used, and decoupling capacitors stabilize the 3 last dynodes. The 
finished printed cards were protected with water resistant varnish and installed 
with bags of dissicant material in the water-tight enclosures mentioned above.

It is important that the gains of the 3 PMTs installed in each detector station 
are matched to each other to avoid unbalances in charge collection. Unmatching 
could impair the 
homogeneity in the response of the detector, and even reduce the dynamic range. 
Because of availability of equipment and space, we used a common high voltage 
supply in each detector station. Because of this, and provided that the observed 
differences in gains were small (less than 15~\%), we compensated the 
differences in gain by using passive, constant-impedance variable attenuators to 
reduce as necessary the output pulse amplitude of the two tubes having larger 
gains in each station.

In order to determine the relative gains we adopted a routine procedure based on 
the measurement of the signal from each PMT produced by background muons. The 
trigger for this measurement is taken from the signals from other PMT belonging 
to the same station, as described in\cite{GAP_PRYKE_CALIB}. Once the average 
relative gains are obtained for the three tubes, the attenuators are set in the 
two PMTs with higher gains, matching the peak position produced by the PMT with 
lowest gain. This procedure for gain matching and calibration has been performed 
on a monthly basis during the complete period of measurements. The system proved 
to be very stable and very few adjustments were required along this time.

In addition to this periodic gain matching monitoring procedure, a daily routine for 
monitoring the overall gain of the 4 detector stations has been performed. It 
also uses the natural background muons falling in the detectors as the source of 
signals for calibration. It has been found that the spectra of the summed 
signals of the three PMTs within each peripheral detector station, and also the 
response of the 3" PMT installed in the small central detector show clearly a 
peak when they are triggered by themselves. Although it is somewhat broad, the 
position of this ``background'' muon peak is very closely the same as the 
position of the similar peak obtained when a pair of external plastic 
scintillators are used to select vertical and central muons for triggering. This is 
valid for both, voltage and charge spectra. This experimental result, which might 
be due 
to the remarkable uniformity in the light distribution produced by the Tyvek 
liners, provides a simple and reliable procedure for remote monitoring and 
calibration of the station gain \cite{gap_00_027}. This peak value 
has been called VEM ({\bf V}ertical {\bf E}quivalent {\bf M}uon), and is defined 
as the charge (or voltage) peak produced by singly charged, energetic particles, 
crossing vertically the detector along its axis. This VEM-value is a characteristic 
parameter 
of each detector, and depends on its components, geometry, construction, and 
also on its operation conditions (transparency of the water radiator, bias 
voltage, etc). The VEM-value provides a practical way to normalize the signals 
from different detectors and moreover, to express the total signal produced in 
each station by an EAS ({\em i.e.} muons, electrons, gamma rays, etc., hitting 
the station) in terms of an ``equivalent reference particle''. Muons have been 
selected in this case as they are present everywhere and proved to be very 
convenient for calibration.

In previous studies\cite{gap_97_032} performed with the prototype 
detector, very good homogeneity in charge collection was obtained by using the 
sum of the signals of the three PMTS. This behavior, which might be again 
attributed to the excellent light spread produced by the Tyvek liners, is kept 
almost independently of the entrance points and directions of the muons. 

According to these results, our design included fast active adder circuits 
installed in the peripheral detectors. The operational amplifier employed (CLC 452) 
works also as the driver for the relatively long RG-213 cable, 
carrying the signals from each detector to the DAQ room. Although their response 
in speed is excellent (we require a 130~MHz bandwidth), these circuits introduce 
a limitation in the dynamic range as the maximum span voltage is less than 2 V. 
In order to reduce the reflection of the pick up noise signal the impedance of 
the cable was matched at the sending end, but this reduces further the available 
amplitude to only 1.4 V. On the other hand, an acceptable signal to noise ratio 
in the DAQ room asks for minimum signal amplitudes for single muons of 
$\sim$~100 mV. These figures, together with the measured RF pick up and the 
signal attenuation produced along the cables, limit the final available dynamic range 
from 1 to 15 muons. Even when this dynamic range is limited, it has been found to
be acceptable because only $\sim$~30\% of the events had to be rejected in the 
off-line data analysis due to electronic saturation in any station.

Because there is only one PMT in the central detector, and the cable length to 
the DAQ room is relatively short ($\sim$~45~m), its anode signal was directly 
sent without summing circuit nor attenuator, and hence without the limitation 
in the dynamic range present in the outer detectors.

\subsection {Trigger System and Data Acquisition}

\begin{figure} 
\centerline{\epsfig{file=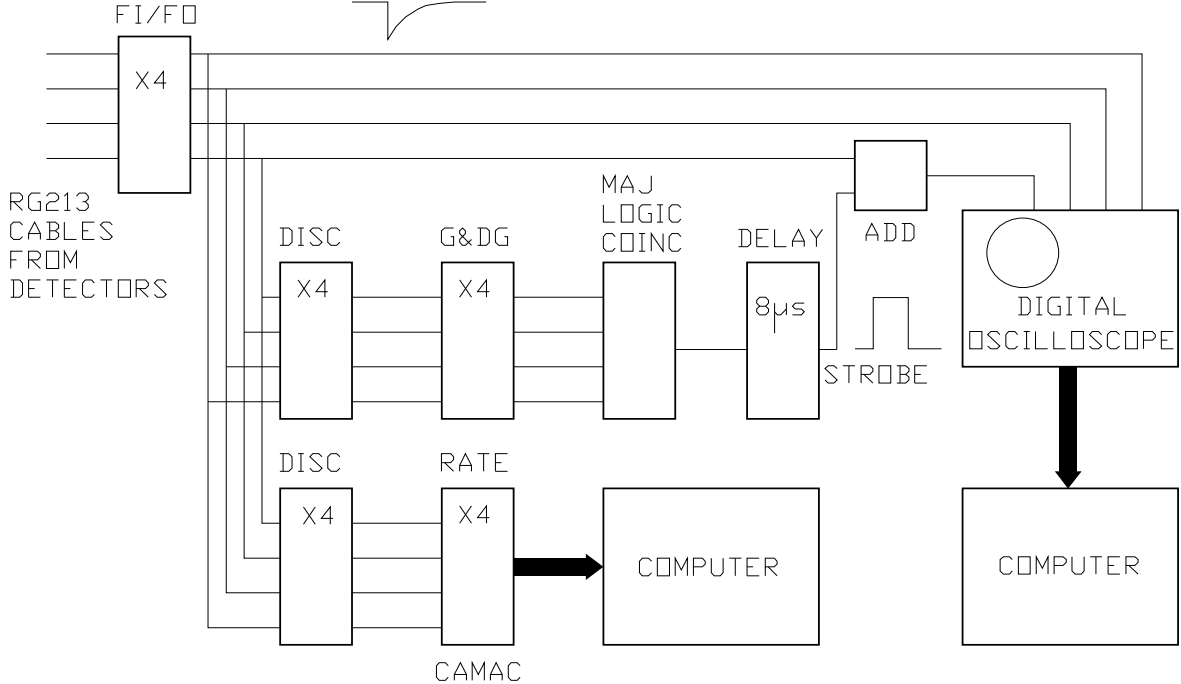,width=16.cm,height=9.5cm}}
\caption {Simplified block scheme of the Trigger and DAQ system.} 
\label{fig:trigger}
\end{figure}

The signals from the four stations arriving to the electronics front panel are 
split using linear fan-in/fan-out (FIFO) modules (see Figure \ref{fig:trigger}).
Then, they are fed directly to 
the input connectors of a four-channel digital oscilloscope (Tektronix TDS 3034 
set at 500 Ms/s). The oscilloscope is the core of our DAQ system and works as 
the digitizing stage for detector signals under control of a STROBE pulse. The 
detector signals are at this point unsynchronized after travelling different 
lengths of cables (206, 196, 310 and 44~m, for detectors A,B,C,D, respectively). 
Thus, in order to generate 
valid trigger conditions, it is essential to compensate these different transit 
times. With this purpose, we use the second signals from the FIFOs to generate 
logical pulses in analog discriminators (discrimination level $\sim$ 1~VEM), 
then these pulses are delayed accurately to compensate for these differences in 
time and then they are fed to a majority logic coincidence unit to select the 
desired trigger condition. The time window in this module is set to ~1.1~$\mu$s, 
covering safely the maximum time used by the EAS front to go across the array, 
even for the case of almost horizontal directions.

\begin{figure}
\centerline{\epsfig{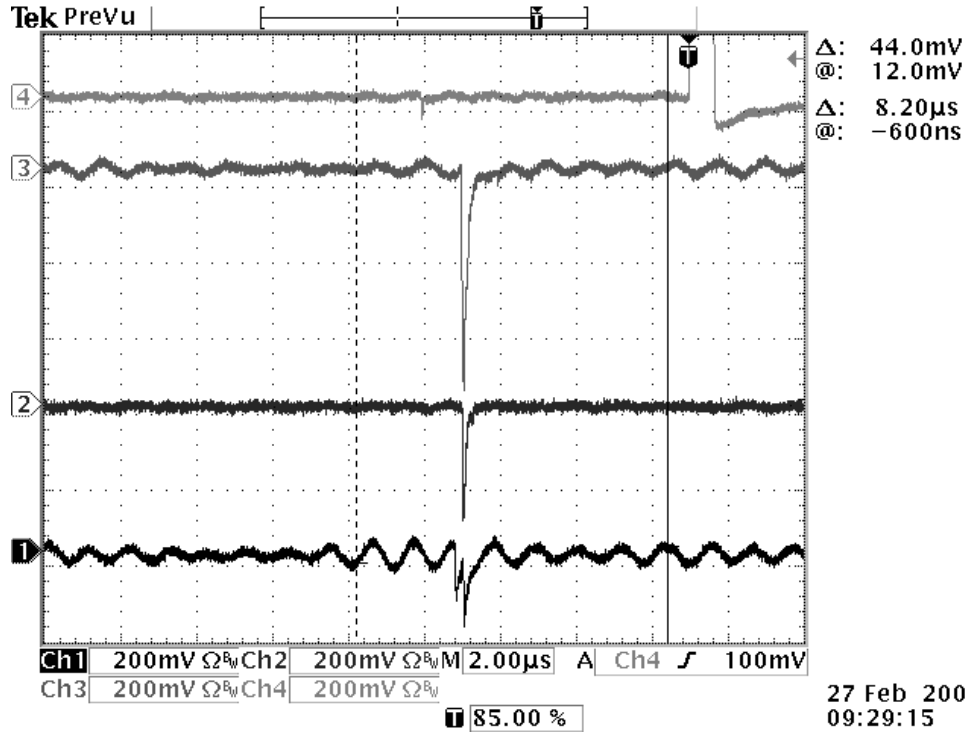}}
\caption {Oscilloscope screen during the capture of a real shower. In 
channel 4 the STROBE signal is summed (with positive polarity) to the signal 
from detector D and indicated with the symbol ``T'' (trigger). In the time region 
between cursors (8 to 16~$\mu$s) the signals from a typical shower can be observed. 
The individual traces has been vertically shifted for clarity.} 
\label{fig:scope}
\end{figure}

The digital oscilloscope available does not feature external trigger. For this
reason one of the analog channels had to be used for triggering purposes, in addition
to its signal digitizing function. With this purpose the STROBE signal generated by the
coincidence unit, indicating the production of an event of interest (in practice 
3 or 4-fold coincidences) is delayed about 8 $\mu$s after arrival of the last detector
signal and then summed to one of the detector channels (channel 4 in Figure 
\ref{fig:trigger} and \ref{fig:scope}). Because of the relatively low singles 
counting rates and with the
introduced delay of 8 $\mu$s no overlaps are produced in practice. Provided that 
the STROBE
pulse is summed with opposite polarity respect to the detector signals the Advanced
Trigger feature of the oscilloscope could be safely used for triggering.

When the STROBE pulse is detected by the oscilloscope, the SAVE procedure 
is initiated, {\em i.e.}, the traces stored in the four channel memories 
corresponding to the last 16384~ns, are frozen and transferred to the PC 
disk. This time slice allows us to obtain a good measurement of both, the 
desired detector signals and the unavoidable radio noise pick-up in the long 
cables carring the signals.

\begin{figure}
\centerline{\epsfig{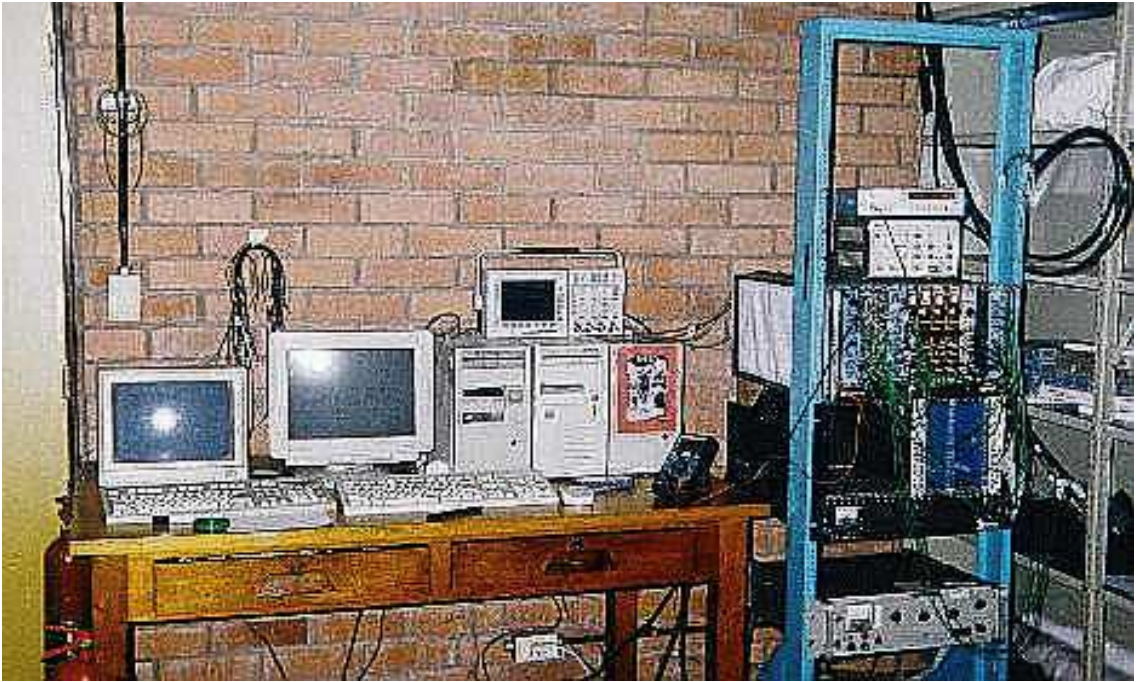}}
\caption {Data Adquisition Room. The core of the DAQ system is the digital oscilloscope.
On the right can be seen the cables connecting the DAQ room with the detectors.} 
\label{fig:setup}
\end{figure}

The system dead time (digitalization, data transfer and PC storage) 
is 22 seconds for event. This relatively long dead time is primarily produced by the 
transfers, through the RS-232 serial port working, at 19200 bps. This dead time
is considered acceptable in comparison with the average time between events, which
is of the order of 6 minutes. 

The first time region of 8192~ns (up to the first cursor in Figure \ref{fig:scope})
is used to compute the bias level at the time of presentation of the detector signals. 
The typical pick 
up noise appears as a dominant oscillation with a period of the order of 
1~$\mu$s, corresponding mainly to the AM broadcasting stations. The following 
8192~ns region, between the cursors, is the time region where the detector 
signals are stored. The last region which contains the STROBE signal is 
not saved to disk.  
The internal 150~MHz bandwidth low-pass filter built in the oscilloscope is 
active in order to reduce the amplitude of higher frequency signals. A fast 
Fourier analysis of the detector signals indicated that their main harmonic 
components extend up to $\sim$~100~MHz, thus little distortion in the 
detector signals is introduced by the filter.

A special program was written to drive the data acquisition system in a 
completely automatic way. Normal collection of shower events is performed when 
the program runs in ``Survey Mode''. The detection of a STROBE pulse causes that
the oscilloscope traces recorded in the 4 channels are saved to disk, together with 
information on the year, day, and local civil time, which allows to reconstruct the 
equatorial or galactic coordinates of the shower arrival direction.

In addition, every day the program switches at a predetermined time to the 
``Calibration Mode''. In this mode, the collection of data from each detector 
station is self-triggered, in order to record background events to calibrate the 
stations, {\em i.e.} to determine the daily VEM value for each station. The four 
detectors are measured sequentially in this mode, under 
program control. The data are stored to disk and analyzed off-line. Roughly, one 
hour and a half is required to acquire 3000 background events (found to be adecuate
to obtain the VEM values with an error of $\sim$~5\%) for each station and to save 
the calibration data from the four detectors. The starting time for the calibration 
procedure, and the total amount of background events for each detector, are set 
in an ASCII file. Once the calibration is completed, the program  automatically 
switches to the ``Survey mode'''' described above. This mode of operation is kept until 
the ``Calibration Mode'' is called up again, at the programmed time next day.

The singles counting rates of the four detector stations are permanently 
recorded using a CAMAC scaler with a refreshing time of 1 s, and are also 
saved to disk. This information is valuable for monitoring the status of each 
station. It helped discarding particular data when the operating 
condition of a particular station became unstable due, for instance, to a high 
level of pick up noise, or gave an alert signal for the need of maintenance of a 
station, in occasional cases of light leaks. The recording of the counting rates 
is also program-controlled and does not require operator action to run, once it 
is launched.

\section{Simulated performance of the array}

In order to characterize the behavior of the array, detailed simulations were 
performed to estimate its efficiency for shower detection and its angular and 
energy resolutions. A special routine, simulating the detector response to the 
different shower particles, has also been written to provide an input for the 
reconstruction routines.

\subsection{Shower database}

The AIRES program \cite{AIRES} using the SYBILL hadronic package was used in the
first step of the simulation pipeline: the construction of an adequate shower
database containing detailed information about the secondary particles at ground 
produced by primary cosmic rays of the energies of interest.

The shower simulation starts with the injection of a primary particle in the high 
atmosphere ($\sim$ 100 km above sea level) and tracks down the 
different generations of secondary particles in the subsequent cascade. The technique 
known as {\em thinning}\cite{AIRES} was used to reduce the CPU time and the disk 
storage requirement. This procedure consists in tracking explicitly all 
particles above certain energy threshold (or {\em thinning} energy), and those particles
having an energy level below the threshold are computed using statistical weight. 
In our simulations we set a relative thinning energy level of 5.10$^{-5}$ with respect to 
the primary particle energy.

\begin{figure} 
\epsfig{figure=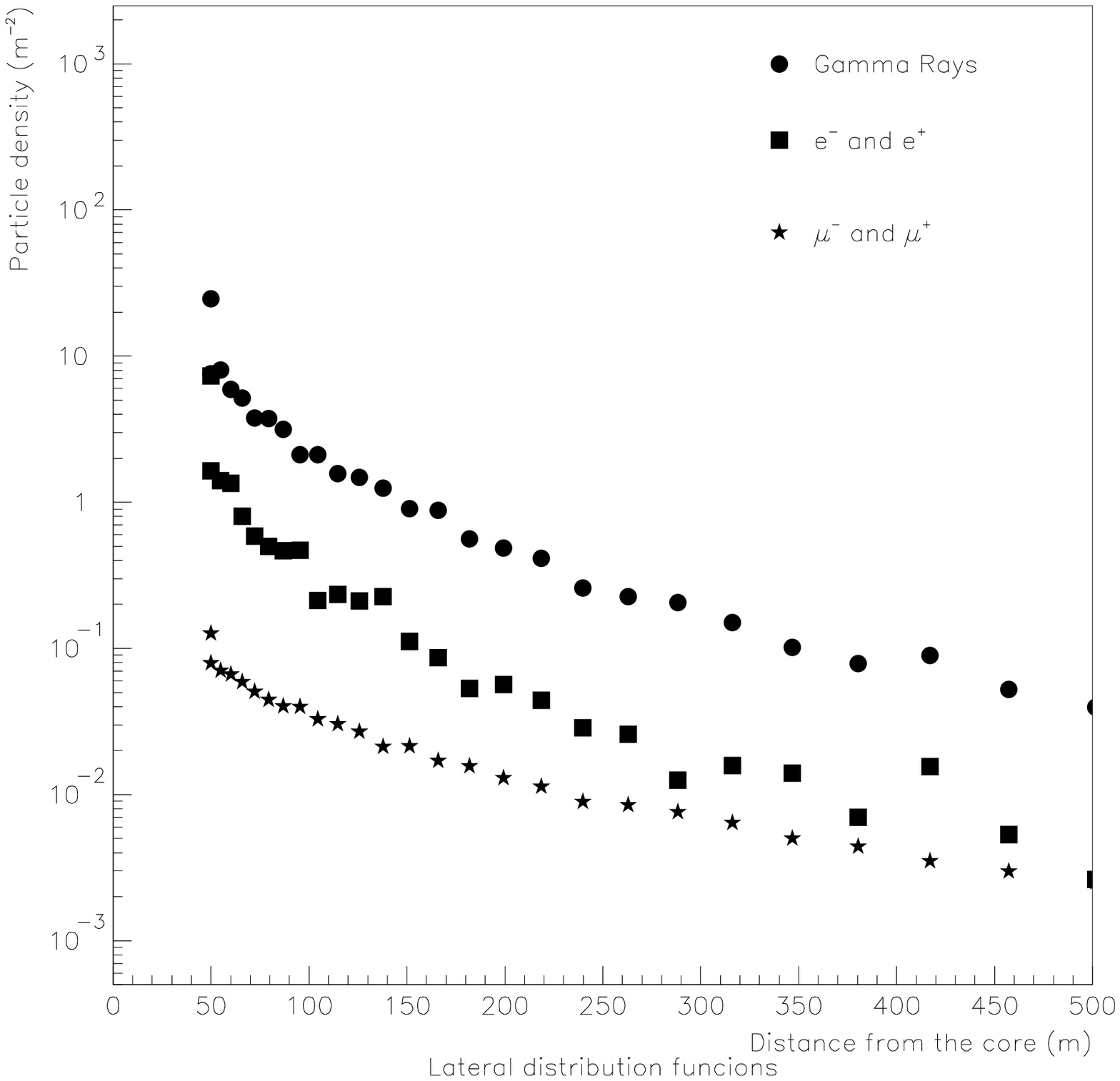,width=7.5 cm,height=7.5 cm}  
\epsfig{figure=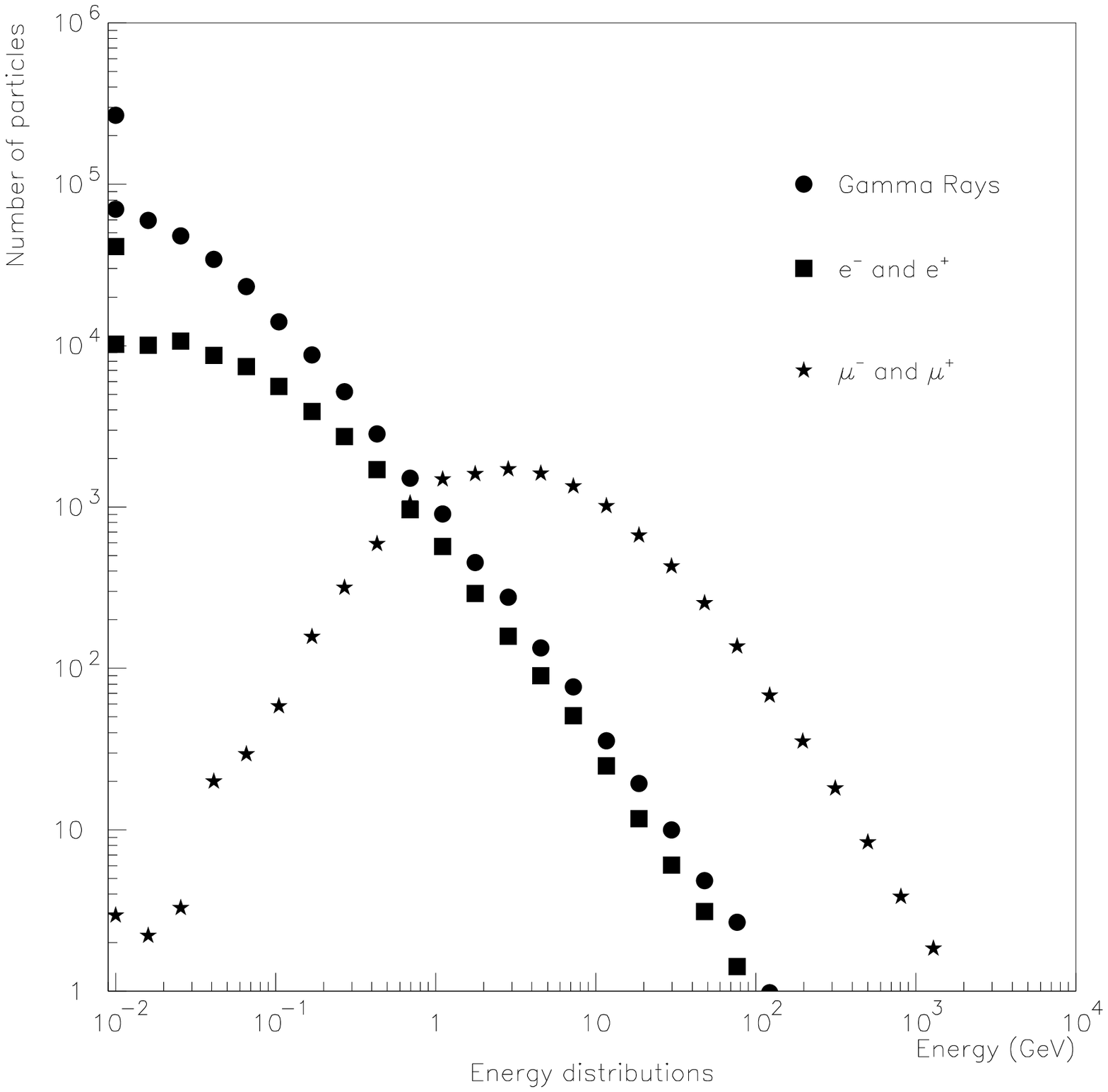,width=7.5 cm,height=7.5cm}
\caption{Lateral distribution function and energy distribution for different particle species
in the cascade. These distributions corresponds to a proton primary of~2.10$^{15}$~eV and 
a zenithal angle of 30\degree. These curves are typical AIRES results.}
\label{fig:aires}
\end{figure}

To construct the shower database, twenty primary energies ranging from 10$^{14}$~eV to 
10$^{18}$~eV were selected, and two nuclear species (protons and iron nuclei) were considered 
as primary particles. They were injected at zenithal angles from 0\degree to 60$^{\circ}$, 
in 15\degree steps. To reduce the artificial fluctuations due in part to the thinning method
\cite{gap_96_020}, and also to obtain representative values of the relevant parameters,
batches of 100 showers were simulated under the same initial conditions (as described above),
and their average and RMS values were used. All these simulations were performed 
considering a ground level of 15 m.a.s.l.

The AIRES program produces a set of tables written in ASCII code, in which the information
referent to the secondary particles reaching ground level (after deconvolution of 
the thinning algorithm) can be expressed simply as particle densities as a function of
the core distance. This function is called {\em lateral distribution function} (LDF). 
In addition, for those particles ``reaching'' the ground level, these tables provide the 
landing time as function of the shower core distance and also their energy distribution. 
These tables include the mean, RMS and extreme values for each computed variable bin.

Figure \ref{fig:aires} shows a typical example of AIRES results for the LDF and for the
energy distributions produced by a proton primary of 2.10$^{15}$~eV impinging at a zenithal 
angle of 30$^{\circ}$.

Thus, by running protons and iron nuclei as primary particles, the AIRES database contains an
amount of 20000 simulated showers covering the energy and zenithal range of interest for the 
TANGO Array. The shower database tables contains only particle densities, energies and
arrival times (with respect to the core position particles arrival times) for muons, electrons,
and gamma-rays.

\subsection{Array simulation procedure}

In order to predict the response of the array, the information on showers contained 
in the AIRES 
database tables was used to simulate "events", {\em i.e.} the effect of individual showers 
falling relatively close to the array. 

A simulated shower is a set of information describing in detail the calculated number and properties of 
secondary particles reaching the ground level. A simulated event is the set of information about the 
calculated effect of the shower on the array, taking into account the simulation of the detector, DAQ 
hardware, electronics, etc. With this purpose the showers are read from the AIRES database tables, 
establishing the number, energy and type of particles hitting each detector station, and their relative 
arrival time. The result of an event is a set of electronic signals from the detectors which are
stored in computer memory. The simulation of each particular event allows to determine if it
triggers or not the DAQ system. A total number of 360000 ``events'' , including all
primary species and energies have been simulated from the information contained in 
the AIRES tables and they constitute the {\em simulated events database}.
The procedure to simulate one event is described as follows:

\begin {itemize}

\item For each primary energy, 9000 shower core positions were selected landing at 
random in a area larger than the geometrical area of the array. In 
this way we can estimate the effect of showers falling outside the boundaries and
obtain an estimation of their triggering efficiency. 
The size of this landing area was scaled logarithmically with the 
primary energy with the purpose to take into account the increase of the 
shower size at ground with energy. 

\item For each core landing position the zenithal and azimuthal angles of the
event were chosen as follows: the azimuthal angle was uniformly distributed, and the 
zenithal angle distribution can be described by a cos$^{3}$($\theta$) function, with a cut-off at 
45~$^{\circ}$. This cut-off was selected accordingly with the atmospheric depth at 
Buenos Aires, where most EASs arrive within a cone of $\sim$~40\degree. The exponent of 
the distribution was chosen so as to produce a distribution flatter than the flattest one 
reported up to date \cite{Milagrito}. This was done with the purpose of including in the 
database a number statistically significant of simulated events at higher zenithal angles.

\item Once the event core position and the angles were established for each particular 
event, the distances from each detector station to the core were calculated. 
Then, from the AIRES tables the apropiate mean values and dispersions were extracted and
interpolated to reproduce the simulated event.

\item To include the shower-to-shower fluctuations, uniform random number generators 
were profiled (using the accept-reject technique) \cite{ac-rej} to reproduce the mean 
value and dispersion of the AIRES particle density tables contained in the database, 
according to the particular secondary particle considered. With the AIRES tables interpolated to the 
particular conditions of each simulated event, and the modified random number generators, 
the densities (particles/m$^{2}$) of muons (both charges), electrons (both charges) and 
gamma-rays hitting each detector neighbourhood, were obtained. Finally, these density 
values were scaled according to the geometrical area of each detector to obtain the 
number of particles falling over each detector in each shower.

\item  The energy and arrival time of every individual particle hitting each 
detector station were obtained using the same procedure (the accept-reject
technique). This was made taking by into account the particle species and its 
distance to the core.

\item Once the number of particles, energies and arrival times of all particle species
falling on each station for the event, were obtained, the detector signal was
obtained as is described in detail in \ref{lab:simdet}.

\item  The next step in this calculation was the simulation of the response of the 
data acquisition electronics by performing a check to determine whether each 
particular simulated shower produces or not a valid trigger. With this purpose,
 the simulated traces 
for each detector station were scanned, searching for the threshold crossing 
times in each channel (there could exist multiple crossings in a single event). 
Then, these threshold crossing times, determined in each channel, were compared 
to those corresponding to the other channels to establish the presence of 
temporal coincidences between the traces (an EAS). If multiple crossing times
were present in one or more channels, each one of them was searched for
time coincidence with the other channels. The threshold levels in all channels 
were set as equivalent to the signal amplitudes produced by 1~VEM from each 
particular detector, and the time window for the coincidences was set 
to~1.1$\mu$s, in correspondence to the real situation during measurements. If a 
coincidence condition was found, then the event was 
classified accordingly to the number of stations involved in the coincidence.

\item Finally, the behavior of the A/D converter stage was also simulated, featuring a FADC 
working at 500 Ms/s (like that used in the data acquisition system). An 
appropriate noise generator has been included. From noise spectrum measurements 
we concluded that the local AM radio stations are the main noise sources, 
contributing with $\sim$~15 to 30~mV to the signal (the typical signal amplitude 
corresponding to one single particle is $\sim$~100~mV). 
The noise spectrum can be described as a continuous distribution with 
superposed, strongly varying peaks corresponding to the well-known local AM 
broadcasting frequencies, ranging from~$\sim$~550 to $\sim$~1650~kHz. The 
FM band is also seen in the noise spectrum. However, its amplitude is much lower 
and can be safely ignored. On this basis, in order to obtain a realistic 
simulation, we added to the simulated signals a noise spectrum which follows the 
description given above.

\end{itemize}

All computer programs required for the simulation pipeline (except AIRES) were 
especially developed in the present work. 

\subsubsection {Detector simulation}
\label{lab:simdet}

A simple and very fast simulation program was written to emulate the detector
response. In this program, instead of simulating in detail the production and
transmission of the \v{C}erenkov photons emitted during the passage of charged particles
through the water, we used the detailed knowledge of the detector behavior achieved during
the previous years of operation of the first prototype. On this basis, the detector
response was reproduced accordingly to a large set of measured parameters. In
the following, prior to describing the detector simulation program, we present a
summary of the experimental data, obtained previously.

\begin{figure} [ht]
\centerline{\epsfig{file=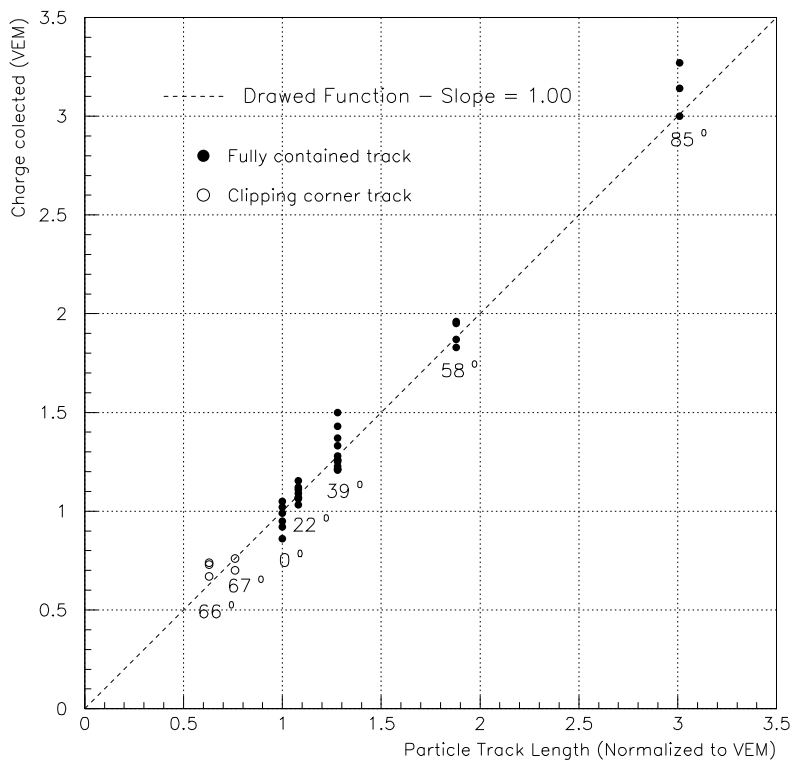, width=11.5cm,height=11.5cm}}
\caption{Sum of the charges collected in the three PMTs as a function of the 
particle track length in the detector water. The muon zenithal angles are also 
indicated. The identity function was drawn for comparison purposes. See 
text for details.}
\label{fig:carla_thesis}
\end{figure}

In previous experiments \cite{nimtank,gap_Carla} the response of the WCD to 
vertical and tilted muons has been observed in detail. In 
these experiments, the entrance and exit points of the muons on the detector 
surface have been carefully selected to cover as much as possible all possible 
situations. A total of 38 different particle track lengths have been measured, 
corresponding to 162 different situations derived from the symmetry properties 
of the detector configuration. A particle track is considered to be ``fully contained'' 
when the entrance point of the muon is anywhere on the lid and the exit point is 
in the bottom, or when the entrance and exit points are near diametrally placed on 
the lateral cylindrical wall of the detector. Either an entrance or exit point on 
the side wall, and the other in the bottom or in the lid, are considered to produce 
a ``clipping corner'' track. As a result of these measurements (which are summarized
in the Figure \ref{fig:carla_thesis}) we have found that the 
sum of the charges collected in the three PMTs of our WCD is, very approximately, 
directly proportional to the track length of the particle in the water radiator, and 
this is valid regardless of the entrance point position or the zenithal angle of the track.

For all measured tracks the digitized pulse shapes were recorded. The rise and fall 
times remain almost constant for the whole range of track lengths, which might be 
understood from the fact that these parameters are primarily determined by the 
highly diffusive properties of the Tyvek liner\cite{black_top}. These measurements 
have also shown that the fluctuations of the measured parameters (rise and fall times, 
voltage amplitude, and charge) are not larger than about $\pm$ 10~\% of their mean 
value. These results were supported by GEANT \cite{geant} simulations, performed
previously\cite{gap_96_011,gap_96_029}. 

In addition to the response to fast muons, the response of the WCD to fast 
electrons and gamma-rays was obtained. Both, electrons and gamma-rays, produce 
also an amount of light proportional to their track lengths. It should be taken 
into account that gamma-rays are detected through their interaction with the 
water going essentially through pair-creation processes. This is the most 
probable case, given the relative cross sections at the typical gamma-ray energies 
present in the EAS. Therefore, the signals produced by gamma-rays are roughly the 
same as those produced by fast electrons, provided their energy distributions
are similar (see Figure \ref{fig:aires}).

It should be taken into account that the signal produced by a muon with energy
higher than~$\sim$~400~MeV becomes indistinguishable from the signal produced
by an electron with energy higher than~$\sim$~250~MeV \cite{pryke_phd}. Hence,
these values were used to normalize the signals from electrons to those 
corresponding to muons.

In order to include in the simulations the effect of the signal distortions in 
the cables, we have recorded in a previous work the average pulse shape for 
vertical muons transmitted through 200~m of RG-213 cable.

By taking into account all this information, the simulation of the 
surface detector signal was carried out as described below:

\begin{figure} 
\epsfig{figure=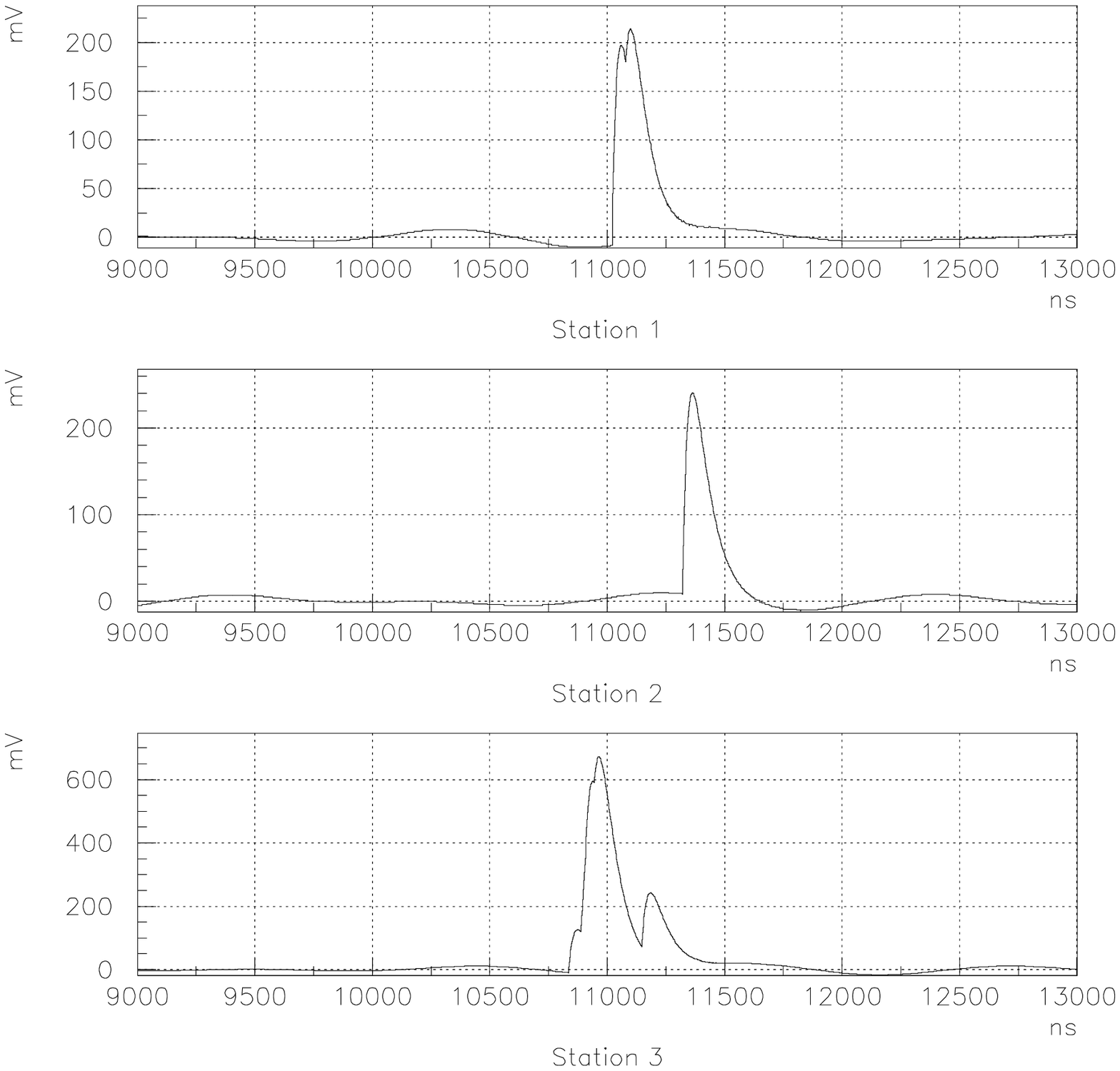,width=7.5 cm,height=7.5 cm}  
\epsfig{figure=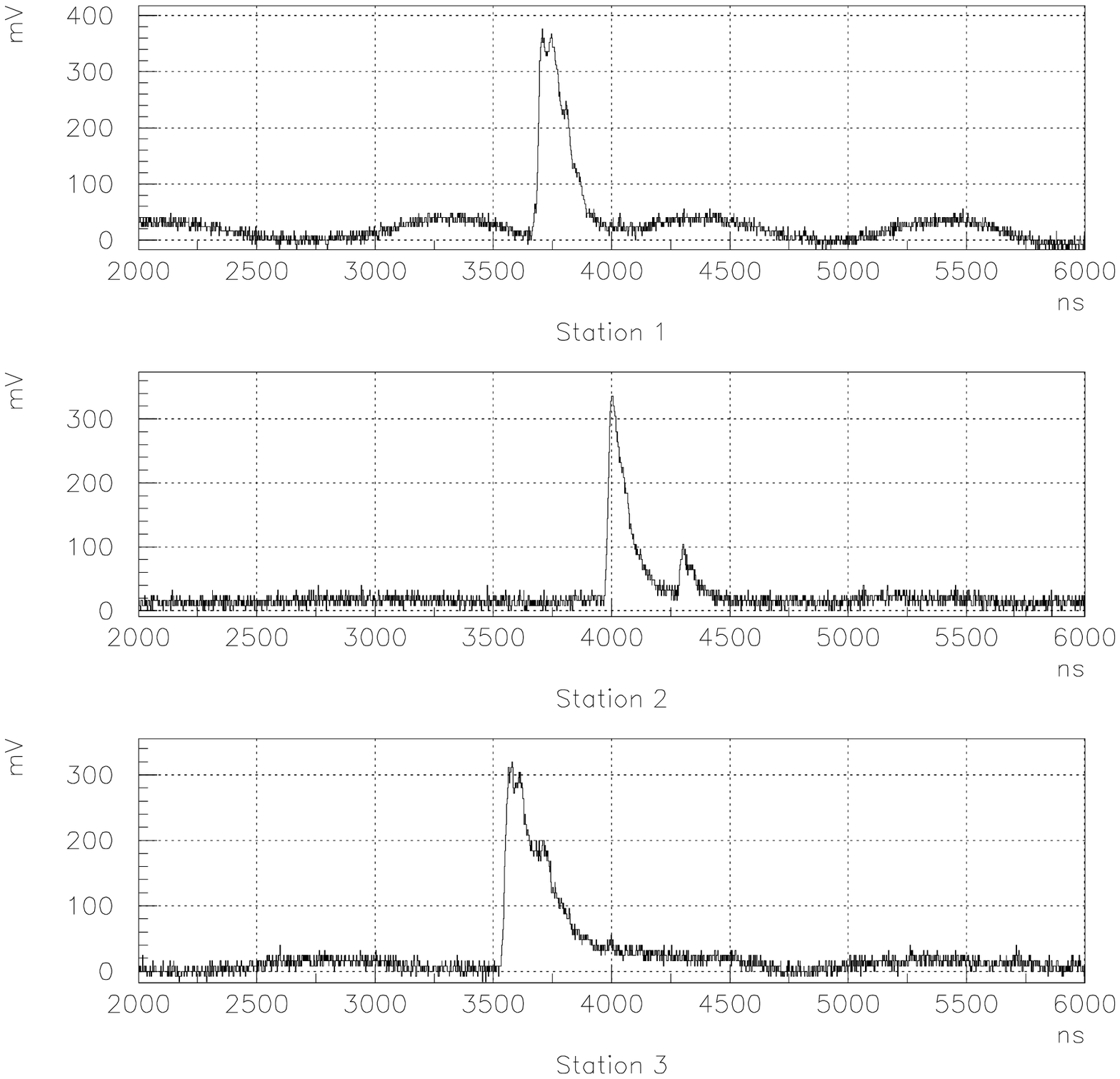,width=7.5 cm,height=7.5cm} 

\caption{Simulated event (left) and pulses recorded from a real coincidence event 
detected in the three large WCDs (right). Both events were chosen arbitrarily 
and are shown only for comparison. The overall zero time for the simulated event is
arbitrary and uncorrelated with the zero time for the measured one. Note the simulated noise 
pick up and the different vertical scales (adjusted automatically by the plotting program).}
\label{fig:sim_vs_data}
\end{figure}

\begin {itemize}

\item {\bf Muons:} For each muon hitting a detector station, a zenithal angle 
is selected using a gaussian-shaped random number generator, with its mean value 
centered in the zenithal angle of the primary particle of the EAS, and a sigma value of 
4$^{\circ}$. Hence the particles are restricted to an angular range of about $\pm$ 
25\degree\cite{pryke_phd}. Once the zenithal angle is established, the range 
of the particle in water is obtained according to its energy, and a peak 
amplitude is found as a function of its range. If the range of the muon exceeds 
the track length inside the surface detector, then the amplitude is made 
proportional to the track length. Finally, rise and fall times are selected with 
a gaussian shaped random number generator and the pulse shape is written to 
memory, considering the respective time delay from the AIRES results (again 
conveniently spread using a gaussian random number generator). The signal peak 
amplitudes, as well as the rise and fall times, are established also from 
gaussian-shaped random number generators, with their relative sigma-values 
obtained from measurements.

\item {\bf Electrons:} The general procedure is similar to that described for 
muons. The main difference occurs in the calculation of the range, which, in the 
case of the electrons, is assumed to be completely contained within the WCD, 
{\em i.e.} no backscattered electrons are simulated. The values of the peak 
amplitudes are obtained from electron simulations performed previously using the 
program GEANT.

\item {\bf Gamma Rays:} The energy of the $\gamma$-rays originated in an EAS range 
from~$\sim$~10~MeV to~$\sim$~100 MeV, and the main interaction channel in 
water goes through the pair creation process. In this energy regime, the mean 
interaction length of gamma-rays in water is about 80~cm. The track length for a 
specific gamma-ray (which depends on the zenithal angle selected as described 
above), determines the probability of creation of an electron-positron pair. In 
this case the electron simulation routine is called with two electrons, having a 
total energy balancing the gamma-ray energy. The energy of the recoiling nucleus 
is neglected.

\end {itemize}

The resulting final program is very fast; once the AIRES tables are locally available
in a 233 MHz PC running under Linux, it simulates an average of 100~events/minute, 
and produces realistic pulse shapes as shown in Figure \ref{fig:sim_vs_data}, where a 
real shower is compared with a simulated one.

As a summary, we have simulated the output of the digitizing electronic stage, 
reproducing the detector signal on the basis of previously known parameters 
relating the underlying physics with the detector response. The effect of the 
cables on the signal shape and the influence of the pick-up noise (only for the AM band),
have been considered.

\section{Shower Reconstruction} 

The reconstruction of the showers aims to find the direction of the shower axis, 
and to make an estimation of the energy of the primary cosmic ray. This reconstruction 
could be made in principle by performing a careful evaluation of a number of parameters 
which are measured from the oscilloscope traces, and from a comparison with the 
results of the simulations. 

The reconstruction procedure is initiated by the obtention of the direction of the shower 
axis by fitting the arrival times to each detector, asuming a flat shower front.
Once the direction is determined, the core position is found through minimization 
of the lateral distribution function using the particle density falling over 
each station. Then,  using Monte Carlo simulations, it is possible to correlate the 
shower primary energy with the particle density measured by the detector stations.

\subsection{Reconstruction of the shower direction}

The reconstruction of the direction is based on the arrival times of the shower front
particles to each detector station in the array. In order to determine the
``trigger time'' in each station, the voltage signal is time-integrated, and the crossing 
times of charge amplitude values equal to 10\%, 50\% and 90\% of the maximum collected 
charge are determined, with the condition that no dynamic range saturation occurs. 
These times are called $t_{10}$, $t_{50}$ and $t_{90}$, respectively.  These parameters 
behave like ''constant fraction discriminators'' crossing times, and they are valuable for
the comparison of the overall time structure of the station's signals when different 
particle densities are measured.

The $t_{10}$ are good indicators of the arrival time of the shower front to the detectors 
and they are used to obtain the shower axis direction, which is coincident with the primary 
cosmic ray arrival direction. On the other hand, the $t_{50}$ and $t_{90}$ are more closely 
related with the time structure and temporal width of the shower than with the shower 
direction, and can be used to estimate the primary mass composition and also
the core distance \cite{linsley,watson}.

\begin{figure}[!h]
  \epsfig{figure=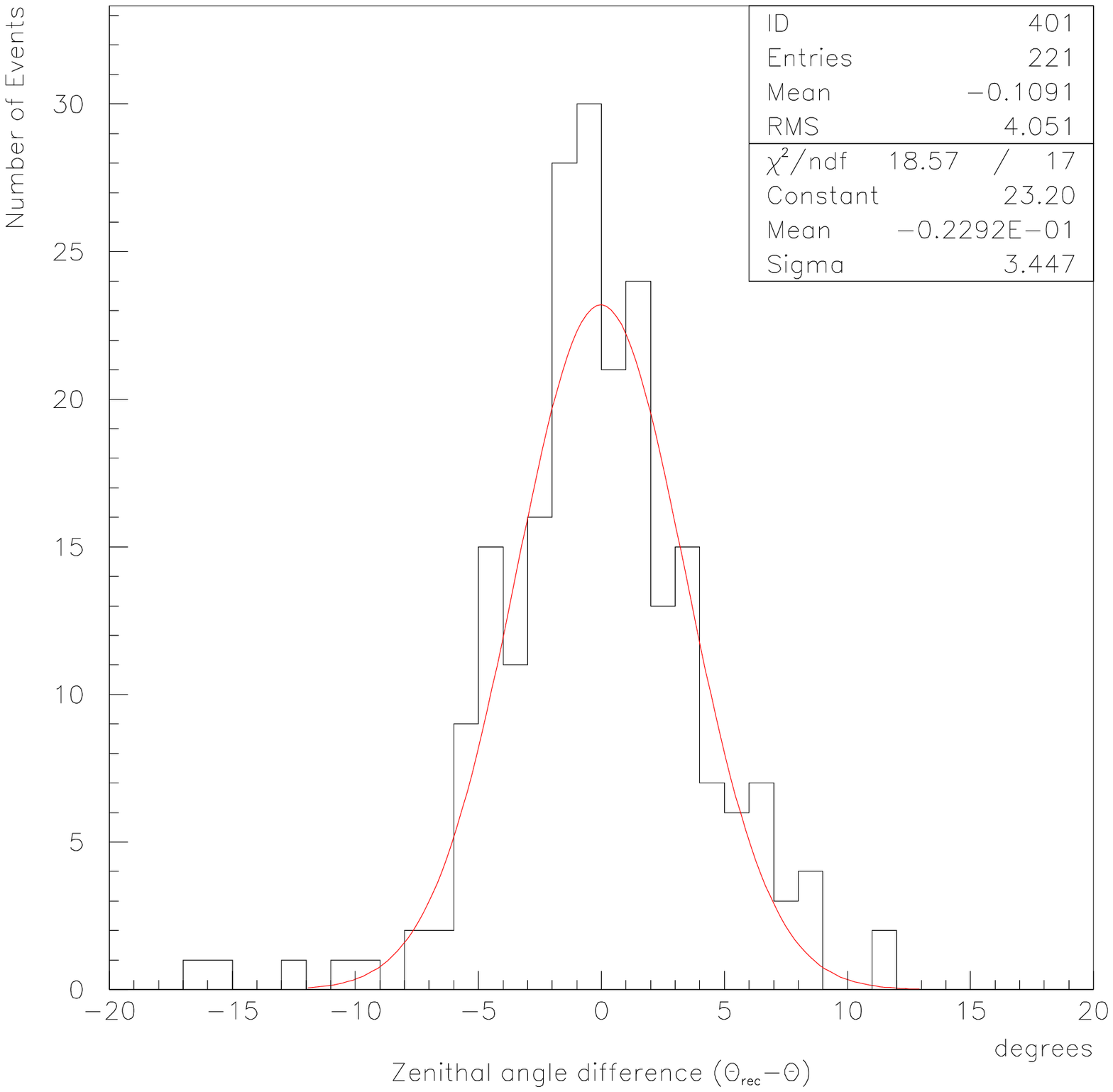,width=7.5 cm,height=7.5 cm}
  \epsfig{figure=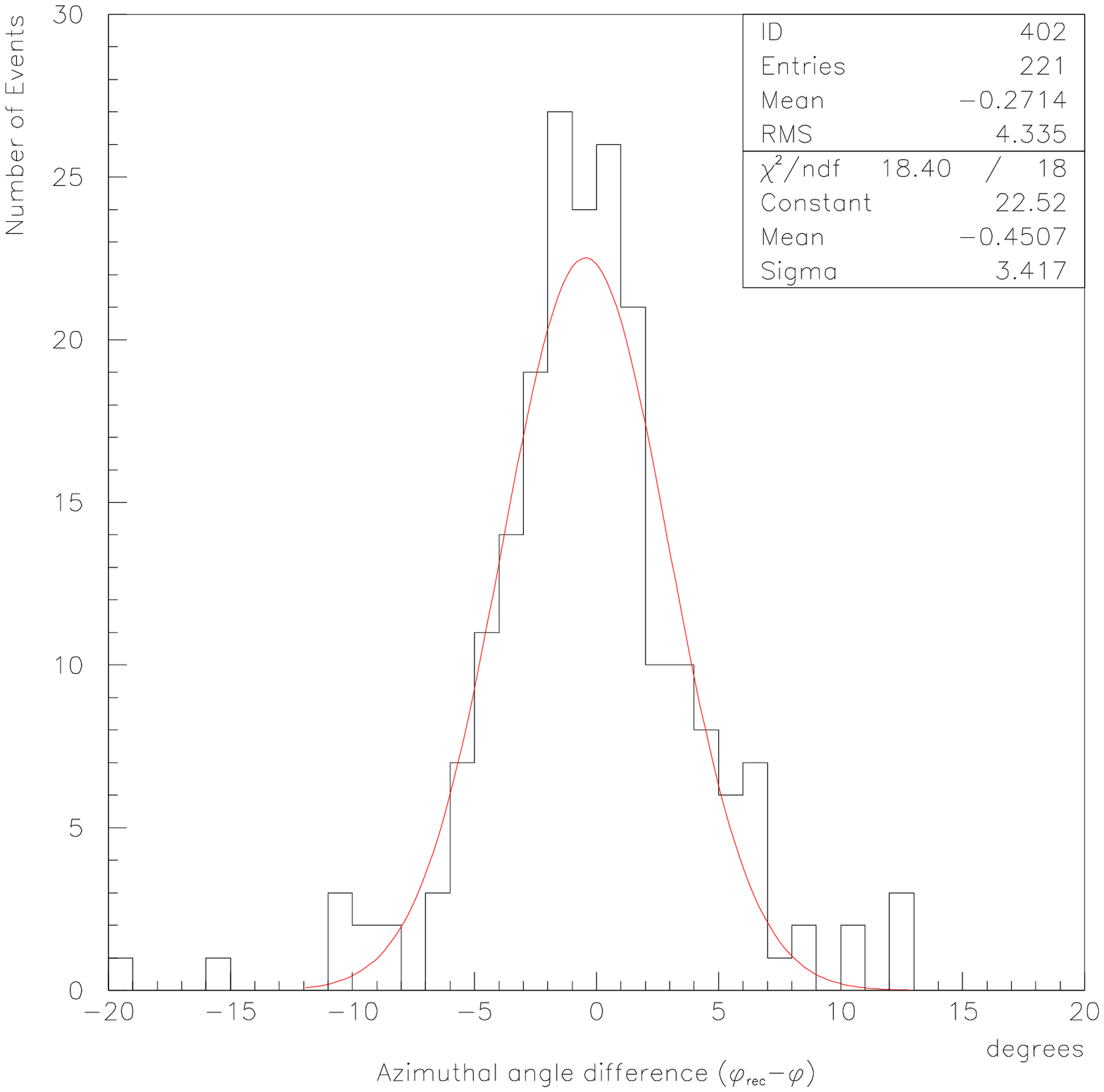,width=7.5 cm,height=7.5cm}
\centerline{\epsfig{figure=ang_res.epsi,width=12.0 cm,height=5.5 cm}}

\caption{Typical gaussian fit to reconstructed zenithal and azimuthal angles. 
The energy dependence of the angular resolution ($\sigma$) is also shown. Filled dots 
correspons to azimutal angle and open dots to zenithal angle.}
 \label{fig:ang_res}
\end{figure}

If three ground stations detect a shower, its axis can be determined by 
triangulation with reference to the arrival times and to the positions of the 
stations. This is made by searching for a unique, downward-going shower front, 
which we assume to be a plane, and moves at the speed of light. When all four 
detectors are hit, then a least squares method is used to find the best fit to 
this plane shower front. More elaborate and detailed algorithms can be used to 
obtain the shower direction including, for instance, the radius of curvature of the 
shower front. However, if we use the available data obtained from our four stations 
and try to use more complicated algorithms they, very often, fail to converge.

\subsubsection{Angular resolution}

As described above, the $t_{10}$ values were obtained from the simulated events database
and used to obtain the arrival direction of each event. From this reconstruction the 
$\theta$ and $\phi$, (zenithal and azimuthal angles) were obtained. These angles are the 
spherical angular components of a vector, normal to the (assumed) plane shower front. 
The accuracy in the reconstruction is determined by comparing these angles with 
the ``true'' angular direction of the particular simulated event, which is read 
from the events database. 

As can be seen in Figure \ref{fig:ang_res}, the angular resolution ($\sigma$)
of the array improves progressively with energy in the decade of 10$^{14}$~eV, 
then remains almost constant in the decade 10$^{15}$~eV and slowly decreases 
beyond $\sim$~10$^{16}$~eV. 

The reconstructed plane is, actually, a plane parallel to the plane tangent
to the shower curved front surface crossing the array at its center point.
Beyond $\sim$~10$^{16}$~eV the shower front disk is much larger than the 
geometrical size of the array and the probability of having the shower core falling 
away from the array, and being still able to produce a trigger, is higher than the
probability for the core to fall closer. Because of the finite radius of curvature 
of the shower front, the vector normal to the tangent plane is more tilted, respect to 
the shower axis, at points lying far away from the core than for points closer to the core. 

Typical shower front curvature radius are in the order of 10 km, hence at a distance of 
300 meters from the geometrical center of the array, the normal to the tangent plane
is tilted $\sim$~2$^{\circ}$ respect to the shower axis. This angular difference 
between the shower axis and the reconstructed direction have to be added to the 
intrinsic angular resolution of the array, which is of the same order of magnitude. 
This effect might explain the decrease in angular resolution at higher energies. Also, 
this may be interpreted as an energy limit for the validity of the flat shower front 
assumption, given the size of our array.

\subsection{Reconstruction of shower energy}

The axial symmetry assumption for an EAS is relevant for the energy analysis. 
It means that in a plane perpendicular to the shower axis, the particle density 
only depends on the radial distance from the axis. On the ground plane this 
symmetry is lost (unless the shower is vertical). However, for moderate zenithal 
angles ($\le$~40$^{\circ}$) the assumption of a symmetric distribution is a valid 
approximation.  For instance, if the EAS has a zenithal angle of 40$^{\circ}$ and the
shower front has a diameter of 300 meters (typical size of the TANGO array) the ''forward''
component of the shower front travels about 250 meters more than the backward'' 
component to reach the ground. Considering the measured attenuation length reported
in the Haverah Park experiment of (780~$\pm$~35)~gr/cm$^{2}$\cite{edge}, the forward
component of the shower front would be attenuated only $\sim$~5\% with respect to the
backward component. This calculation shows the validity of the approximation.
On this basis, we assume in the following that axial symmetry is a valid assumption.

A key parameter required to estimate the energy of an EAS is the LDF, {\em i.e.} 
the particle density as a function of the distance to the core position. From the 
results of previous experiments \cite{HP,AGASA} it is possible to propose the 
functional dependence:

\begin{equation}   
\label{eq:ldf}
{\rho}={A\over{r^{\eta + r/r_0}}}
\end{equation}

\noindent
where $\rho$ is the particle density ([VEM/m$^{2}$]), $r$ is the distance 
to the core ([m]), $A$ is a normalization constant (proportional to the primary particle 
energy) and $\eta$ and $r_0$ control the shape of the LDF. The last two parameters
were obtained by fitting the previous expresion to simulated particle density distributions,
which have included the detector response to different shower particle species ($\mu^{\pm}$,
$e^{\pm}$ and $\gamma$-rays) as described in \ref{lab:simdet}.

\begin{figure}[ht]
\centerline{\epsfig{figure=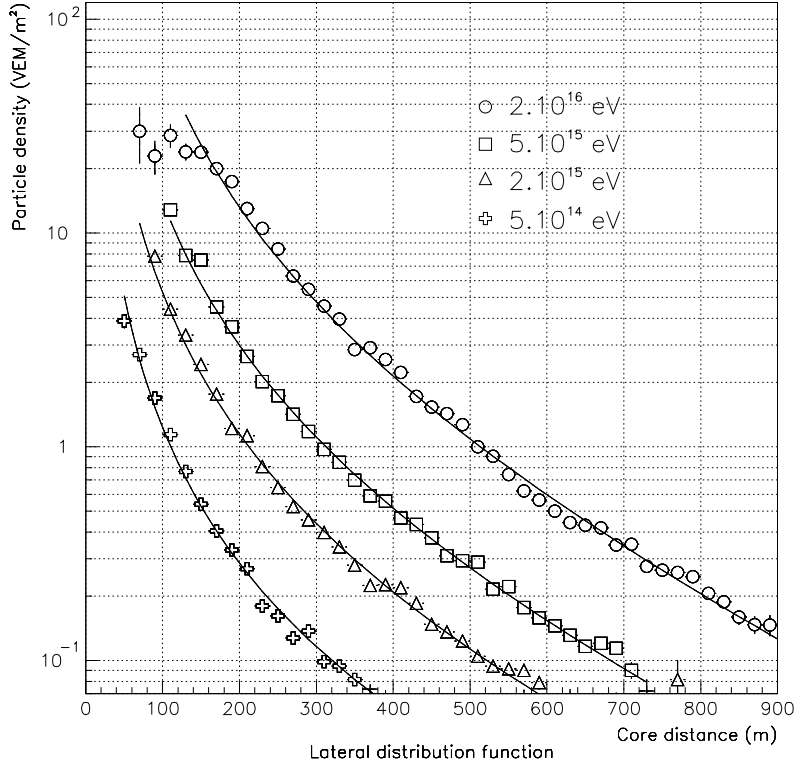,width=11.5 cm,height=11.5 cm}}
\caption{Fits to the averaged simulated lateral distribution function 
for different proton primary energies. The zenithal angles has been averaged in each 
core distance bin.}
\label{fig:sim_ldf}
\end{figure}

Figure \ref{fig:sim_ldf} shows the fits using Equation \ref{eq:ldf} to the simulated 
particle densities ``measured'' with the simulated WCD for several primary energies, where 
all zenithal angles included in the simulation were averaged within each 
core-distance bin. The small ``plateau'' observed in the leftmost part of the 2.10$^{16}$~eV
curve is produced by the (simulated) electronic dynamic range saturation.

\begin{table}[b]
\begin{center}
\begin{tabular}{||c|c|c||} 
\hline \hline                           
	&$\eta$                 & $r_0$ \\  \hline
Proton  &1.99 $\pm$ 0.02        & 3400 $\pm$ 150 \\     
Iron    &1.94 $\pm$ 0.02        & 3400 $\pm$ 100 \\
Average & 1.965                 & 3400 \\
\hline \hline
\end{tabular}
\caption{Lateral distribution function parameters obtained from the simulated events 
database for both primary species. Also, the average values used in the 
reconstruction algorithm are shown.}
\label{tab:parameters}
\end{center}
\end{table}

It should be noted that the $\eta$ parameter is slightly sensitive to the primary particle 
mass \cite{PAP86} as was found by fitting the previous expresion to the 
simulated data events. The reproduction of such dependence from simulations is an 
encouraging result. Although different approaches were attempted to obtain, at least,
a primary mass indicator from the reconstructed events in the simulated database, none
of them were satisfactory, probably due to the simulated shower-to-shower fluctuations
that might mask the small differences in the $\eta$ parameter for different primary
species. Because of this we used in the following an average value for the $\eta$ parameter
between the values corresponding to proton and iron primaries (see Table \ref{tab:parameters}).

It is known from extensive Monte Carlo simulations\cite{HP} that there exists a certain 
distance from the shower core for which the particle density of an EAS correlates with 
its primary energy and also their fluctuations are minimized. In the present experiment 
performed with only 4 detectors, we have used a simplified model where the normalization 
constant $A$ of LDF was correlated with the primary energy instead of the particle density 
at a fixed distance of the core position. The LDF was obtained from particle density 
measurements in each detector station, far away from the core.

The normalization constant of the LDF is found through minimization of the
following equation

\begin{equation}
\label{eq:minimization}
{\chi^{2}}={\sum_{i=1}^n \left(\rho_i-{A\over{r_i^{\eta + r_i/r_0}}}\right) }^{2}  
\end{equation}

\noindent
where $\rho_i$ and $r_i$ are the particle density and the distance between the core
impact position and the {\em i}-th station, respectively, and $\eta$ and $r_0$ 
were obtained from simulations as mentioned before. Finally, the particle density 
measured by each station is obtained by the ratio of the time-integrated
oscilloscope trace (bias subtracted) and the VEM value corresponding to that particular 
detector station. This ratio yields the number of equivalent particles falling in 
the station. Then, the equivalent particle densities (VEM/m$^{2}$) is obtained by simple 
normalization to the respective detector area.

The miniminization of Equation \ref{eq:minimization} was performed through a grid search 
on the simulated data of the events database, yielding the $x$ and $y$ coordinates of 
the core position, as well as the normalization constant $A$. Table \ref{tab:core} 
shows the accuracy of the core position reconstruction obtained by this method for some 
energies. The accuracy is degraded at higher energies, probably because the shower front 
size becomes comparable with the array size.

\begin{table}[b]
\begin{center}
\begin{tabular}{||c|c||} 
\hline \hline                           
	Primary Energy          &  Core position accuracy  \\  \hline
	5.10$^{14}$ eV       &  40 m  \\     
	2.10$^{15}$ eV       &  30 m  \\  
	5.10$^{15}$ eV       &  55 m  \\  
	2.10$^{16}$ eV       & 110 m  \\  
\hline \hline

\end{tabular}
\caption{Accuracy (RMS) of the reconstructed core position.}
\label{tab:core}
\end{center}
\end{table}

\subsubsection{Correction by atmospheric attenuation} 

The axial symmetry asumption proved to be a valid approximation regarding the
forward 'and backward 'components of the shower front for non-vertical showers. However, 
the effect of atmospheric attenuation in a tilted EAS development cannot be ignored. 
In order to get an evaluation of the magnitude of this effect we used the simulated 
events database to estimate the atmospheric attenuation on the shower propagation
through the atmosphere.

By simple geometrical considerations it is possible to propose a functional dependence 
of the form 

\begin{equation}
\label{eq:zen_corr}
{N} = {A} e^{\left[\beta (sec(\theta)-1)\right]}
\end{equation}

\noindent
where $N$ is a normalization factor, proportional to the primary particle energy 
that includes the atmospheric attenuation correction factor and this constant includes 
the atmospheric attenuation correction factor.

For each primary energy, the simulated showers were divided in zenithal angle bins of 
5\degree each, and a fit was performed to the data using the functional dependence 
shown in Equation~\ref{eq:zen_corr}, {\em i.e.}, assuming only a dependence for the 
$A$ parameter on the zenithal angle. The average value obtained for $\beta$ by fitting 
the simulated data to Equation~\ref{eq:zen_corr} is $\beta = 4.1 \pm 0.1$. For the
zenithal range of interest ($\theta \le$ 30$^{\circ}$) this correction because of 
the atmospheric attenuation increases the estimated EAS's energy up to $\sim$~50\%.

\begin{figure}
\centerline{\epsfig{figure=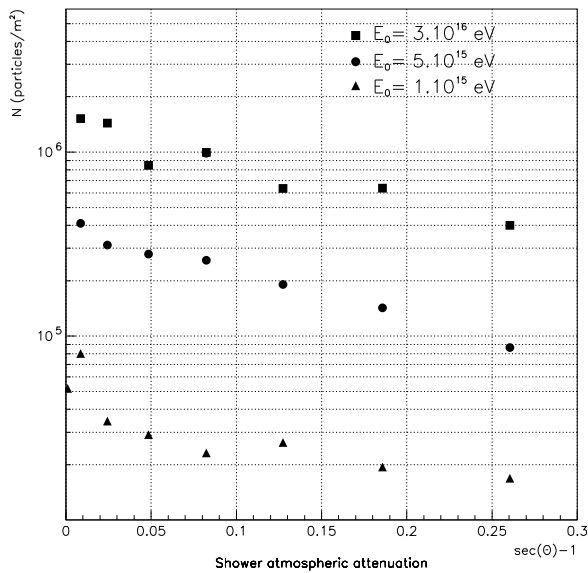,width=8.5 cm,height=8.5 cm}}
\caption{Simulated zenithal atmospheric attenuation for different primary energy.}
\end{figure}

\subsubsection{Primary energy assignment}

Finally, after minimization of Equation \ref{eq:minimization} and being performed the 
atmospheric attenuation correction (for which the directional reconstruction is required) 
it is possible to show the relationship between $N$ -a parameter obtained from the shower 
reconstruction routine- and the primary energy (obtained from the simulated events 
database). It should be noted that in this survey over the simulated events database 
we found that, beyond $\sim$~2.10$^{16}$ eV, $N$ fails to converge, and the linearity 
(in logarithmic scale) as a function of the primary particle energy is lost. 
Therefore, only data at lower energies are shown in Figure \ref{fig:prim_en}.    

From these fits we obtain the following expressions, useful to correlate 
the parameter $N$ [VEM/m$^{2}$] with the primary energy [eV]:

\begin{figure}[!h]
  \epsfig{figure=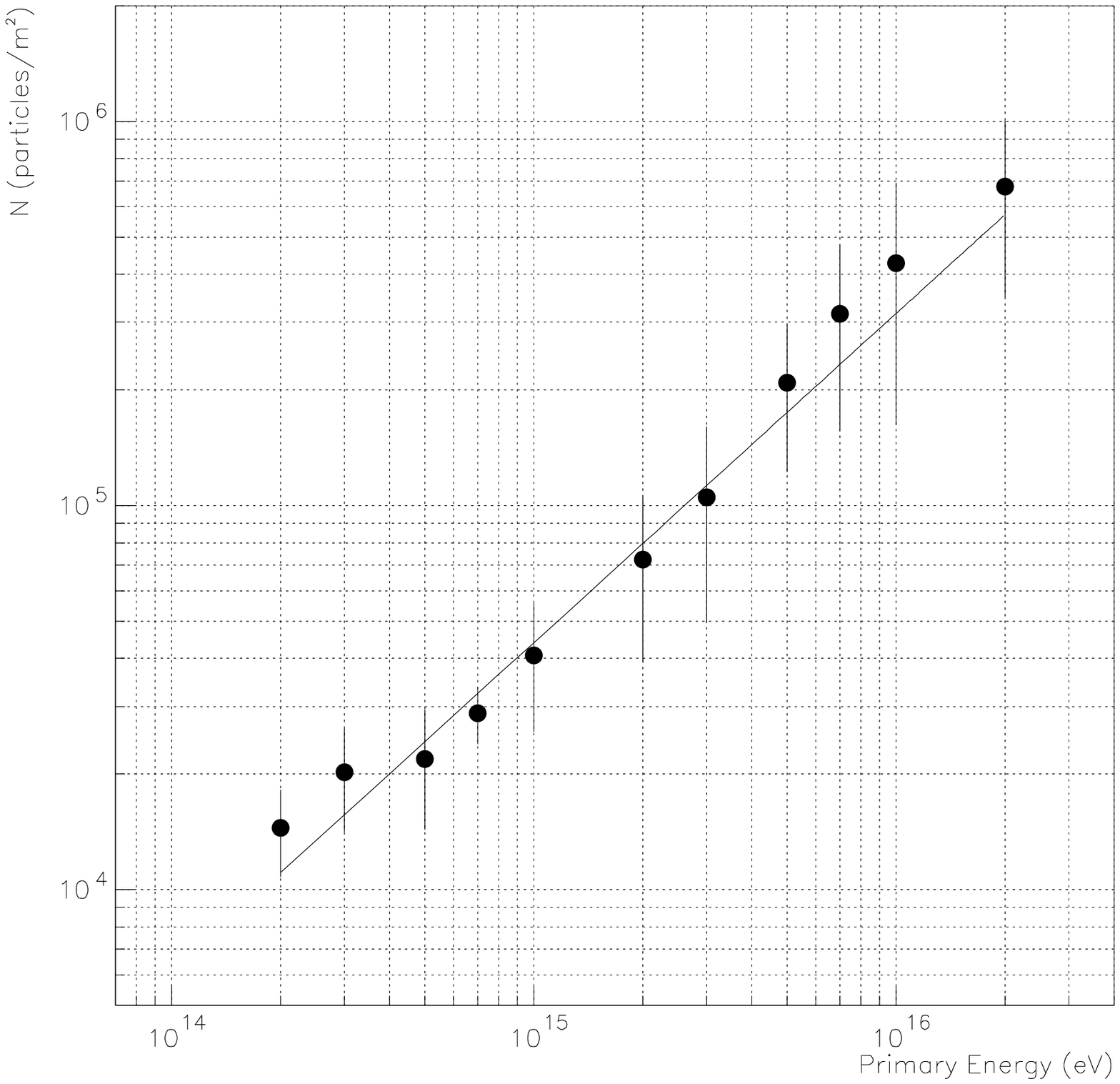,width=7.5 cm,height=7.5 cm}
  \epsfig{figure=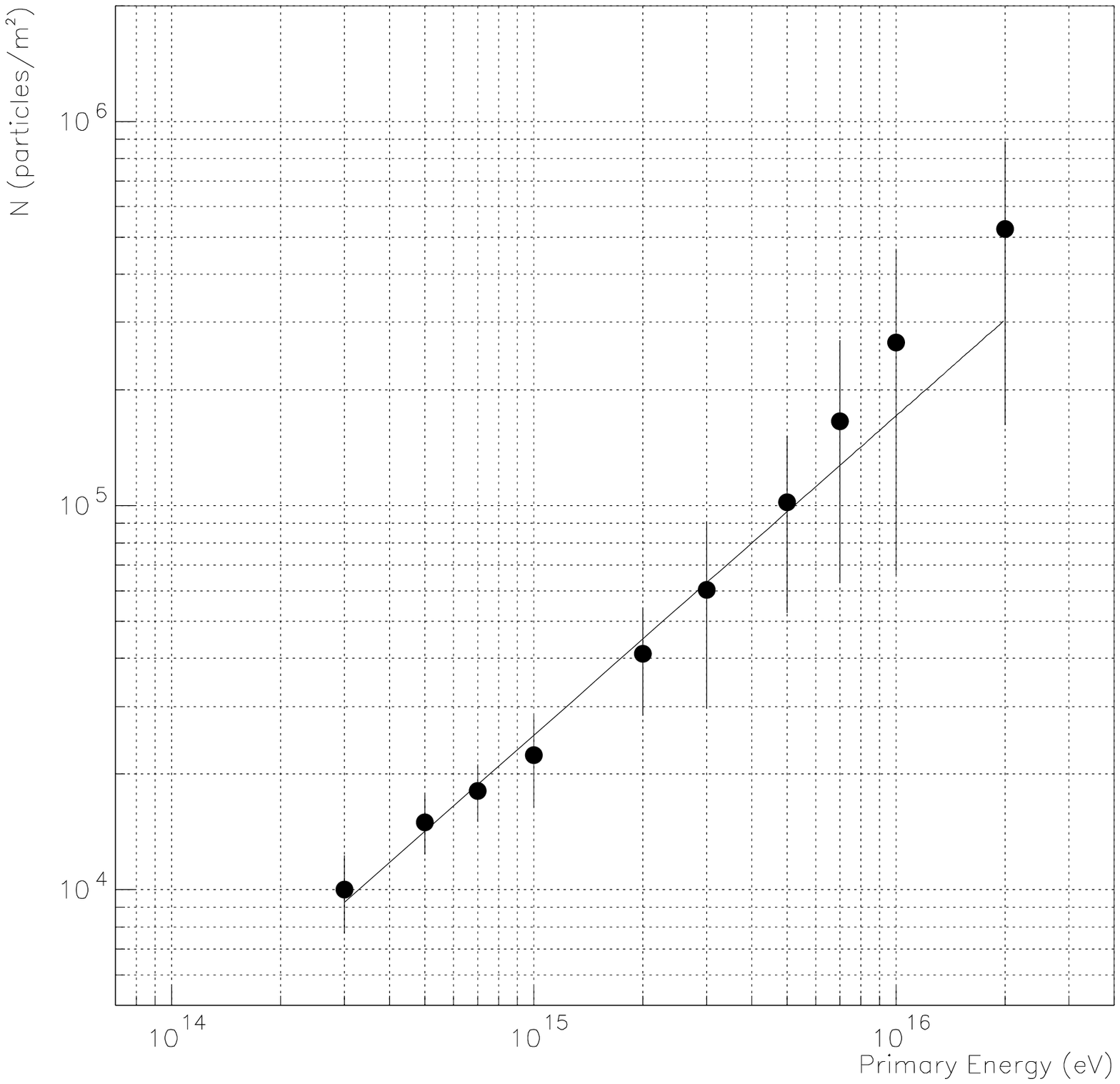,width=7.5 cm,height=7.5cm}
\caption{Relationship between $N$  and primary energy. 
Left: proton primary. Right: iron primary.}
\label{fig:prim_en}
\end{figure}

\begin{equation}
\label{eq:energy_proton}
{E_0} = {(4 \pm 1)  10^{9}  N ^{1.17 \pm 0.03}}
\end{equation}
\noindent
and

\begin{equation}
\label{eq:energy_iron}
{E_0} = {(5 \pm 2)  10^{9} N ^{1.20 \pm 0.03}}
\end{equation}

\noindent
where Equations \ref{eq:energy_proton} and \ref{eq:energy_iron} correspond 
to proton and iron primaries, respectively.

From these equations it is possible to estimate the relative error in the energy 
reconstruction. Even when the expression has a dependence on the $N$ 
value, its dependence is logarithmic, and the $\Delta$$N/N$ value was found to be 
$\sim$ 0.4 from the simulated database. This yields a relative error of 57\% and 
66\% for protons and iron nuclei, respectively, in the energy range from 
~$\sim$~10$^{14}$~eV to~$\sim$~10$^{16}$~eV. 

According to these results, a knowledge of the primary particle mass would be required
to correctly correlate the N parameter with the primary particle energy by choosing 
the proper expresion. Strictly, this fact prevents us to make an unambiguous assignment 
of the primary energy. Furthermore, it should be recalled that both Equations
\ref{eq:energy_proton} and \ref{eq:energy_iron}, were obtained from surveys 
performed on the Monte Carlo simulations, which are dependent of the particular 
hadronic package utilized. On the other hand, however, the results obtained from both 
expresions are consistent within errors.   


\section {Summary}

A new, Extended Air Shower Array has been constructed in Buenos Aires during 1999 
and was commissioned in 2000. It consists of 4 Water \v{C}erenkov Detectors, 
three of them are arranged in a triangular shape and the fourth is near the
center of the triangle. The enclosed area is~$\sim$~30.000~m$^{2}$. 
The detectors placed in the vertices of the triangle have a footprint area of 
10~m$^{2}$, the central detector has 0.5~m$^{2}$. 

Detailed Monte Carlo simulations of the showers were performed using the AIRES 
code with the SYBILL hadronic package. Various computer programs and routines 
were developed to simulate the array response including the surface detector,
front end electronics, pick-up noise, and triggering. It should be noted that
an effort was made to use experimental data whenever possible. The simulated events
database contains a total of 360000 events.

A reconstruction routine has been developed from the simulated shower database. 
According to the simulations, the angular reconstruction resolution is better 
than 5\degree in the range 5.10$^{14}$~eV to 10$^{17}$eV. The accuracy expected in 
the energy resolution is roughly 60\% in the range~$\sim$~10$^{14}$~eV 
to~$\sim$~10$^{16}$~eV. With respect to the primary mass determination it is 
concluded from the present simulations that no unambiguous assignement can be 
made, at present, from the showers measured with our array.

A fully automatic system for calibration, monitoring and data acquisition has 
been built using standard NIM and CAMAC modules and a 4-channel digital 
oscilloscope connected to standard PCs. Data have been continuously collected 
since September, 2000 and the shower reconstruction analysis will be published 
in an forthcoming paper.

\newpage
\section {Acknowledgements}

We are very especially indebted to the late J.Vidall\'e for his unvaluable 
dedication and help in the early stages of the TANGO Array. 

We are also deeply indebted to D.Simoncelli and E.Fisher for their outstanding
work at the Mechanical Workshop in TANDAR Laboratory. Also, we would like to 
express our gratitude to P. Stoliar, H. Di Paolo, C.Bola\~nos, J.Fern\'andez 
V\'asquez and O.Romanelli, for their help with different aspects of the 
electronic system. 
Thanks are given to M.Figueroa and M.Wagner for their work in the 
characterization of the PMTs. We would like to mention H.Grahmann, O.Ruiz, 
E.Altmann, A.Ferrero and A.Etchegoyen. They helped us in many different ways. 
We also thank Prof. Ma Yu Quian, from Beijing University, for the donation 
of the 3-inch PMT used in the central detector and to {\em Pl\'asticos Industriales S.A.}, 
specially E.Carricondo and P.Martelli for the donation of the fiberglass-reinforced tank. 
Finally, we would like to thank Fermilab for the loan of some electronic modules used 
in the experiment (Fermilab Loan C96082).

The work of P. Bauleo, C. Bonifazi and A. Reguera was supported by different 
CNEA fellowships. This work was partially supported by a CONICET grant 
(PIP~4446/96).

\end{document}